\documentclass[sigplan,nonacm,screen]{acmart}
\usepackage{ifthen}
\usepackage{xcolor}
\newboolean{showcomments}
\setboolean{showcomments}{true}

\makeatletter
\newcommand{\mynote}[3]{%
  \ifthenelse{\boolean{showcomments}}{%
   \fbox{\bfseries\sffamily\scriptsize#1}%
   {\small$\blacktriangleright$\textsf{\emph{\color{#3}{#2}}}$\blacktriangleleft$}}%
  {%
   \@bsphack
   \@esphack
  }%
}
\makeatother

\definecolor{asparagus}{rgb}{0.53, 0.66, 0.42}

\newcommand{\cnot}{\texttt{CNOT}}

\newcommand{\sqiswap}{$\sqrt{\texttt{iSWAP}}$}
\newcommand{\niswap}[1]{$\sqrt[#1]{\text{\texttt{iSWAP}}}$}
\newcommand{\iswap}{\texttt{iSWAP}}
\newcommand{\sw}{\texttt{SWAP}}

\renewcommand\footnotetextcopyrightpermission[1]{}

\usepackage{algorithm}
\usepackage{algpseudocode}
\usepackage{dblfloatfix}

\usepackage{amsmath}

\usepackage{amssymb}
\usepackage{siunitx}
\AtBeginDocument{%
  }

\setcopyright{none}
\copyrightyear{2026}
\acmYear{}
\acmDOI{}

\begin{document}

\title{Fidelity-Aware Frequency Allocation and Transpilation Co-Design for Tunable Coupler Quantum Systems}


\author{Dylan VanAllen}
\affiliation{\institution{Syracuse University}\city{Syracuse}\state{New York}\country{USA}}
\email{djvanall@syr.edu}

\author{Evan McKinney}
\affiliation{\institution{Yale University}\city{New Haven}\state{Connecticut}\country{USA}}

\author{Israa G. Yusuf}
\affiliation{\institution{University of Pittsburgh}\city{Pittsburgh}\state{Pennsylvania}\country{USA}}

\author{Girgis Falstin}
\affiliation{\institution{University of Pittsburgh}\city{Pittsburgh}\state{Pennsylvania}\country{USA}}

\author{Gaurav Agarwal}
\affiliation{\institution{Yale University}\city{New Haven}\state{Connecticut}\country{USA}}

\author{Jason Pollack}
\affiliation{\institution{Syracuse University}\city{Syracuse}\state{New York}\country{USA}}

\author{Michael Hatridge}
\affiliation{\institution{Yale University}\city{New Haven}\state{Connecticut}\country{USA}}

\author{Alex K. Jones}
\affiliation{\institution{Syracuse University}\city{Syracuse}\state{New York}\country{USA}}

\renewcommand{\shortauthors}{VanAllen et al.}

\begin{abstract}
Frequency crowding is a fundamental limitation in superconducting quantum architectures, particularly in tunable-coupler systems. We present a framework that explicitly models both coherent spectator-induced errors and incoherent lifetime effects through an error budgeting approach. Using this model, we analyze how frequency crowding impacts gate fidelity as module size and connectivity scale, and formulate a constrained optimization problem to assign qubit and coupler frequencies under realistic separation and hardware constraints. We demonstrate scalable frequency allocation strategies that minimize spectator-induced errors. We further show that increasing qubit count and coupling density within a module leads to a fidelity–connectivity tradeoff. To explore the benefits at the system scale, we have developed a noise-aware transpilation approach called FINESSE, which minimizes error by selecting high-fidelity paths that satisfy connectivity via \sw{} insertion while jointly optimizing downstream gate execution. We demonstrate this physics-informed architecture--transpilation co-design approach for a SNAIL-based third-order coupler that natively realizes the \sqiswap{} 
basis with frequency aware gate fidelities. On SNAIL architectures, FINESSE achieves an average 8.9\% reduction in log-infidelity cost and 6.8\% reduction in circuit depth vs.\ SABRE. We also compare results on IBM Brisbane's architecture.
\end{abstract}




\maketitle

\section{Introduction}
\label{sec:Introduction}
\begin{figure}[tbp]
    \centering
    \includegraphics[width=\columnwidth]{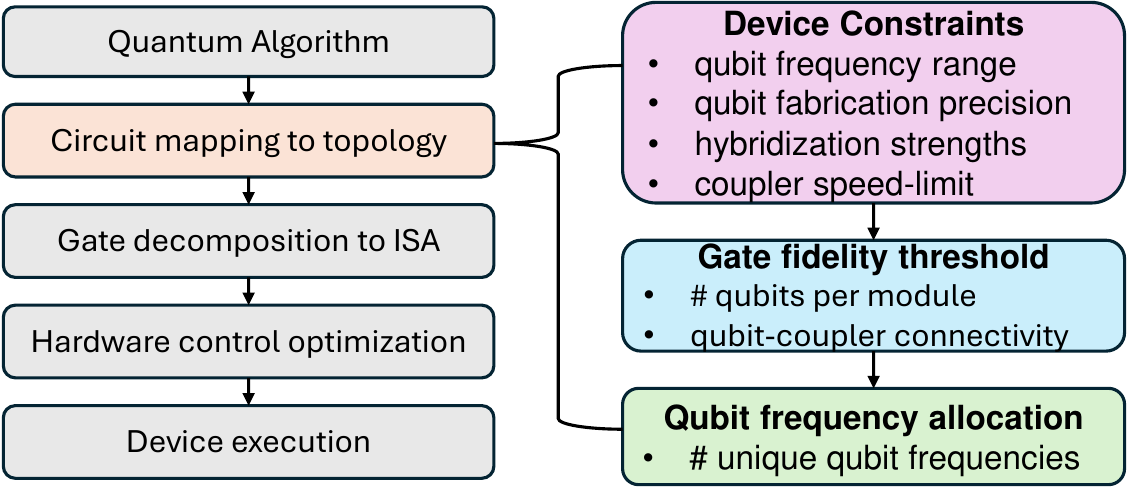}
    \caption{High-level workflow for the co-design of quantum computer architectures~\cite{jones_layered_2012}.}
    \label{fig:codesign-flow}
\end{figure}

Quantum computing holds the promise of polynomial or exponential speedups for problems in optimization, cryptography, and quantum simulation, yet near-term devices remain severely constrained by noise. Recent advancements in superconducting quantum systems have moved toward parametric couplers, which provide useful tradeoffs between quantum gate capabilities and qubit coupling topologies~\cite{mckinney_co-designed_2023}. These developments highlight the need for a physics--system co-design approach that jointly considers hardware constraints and compilation methodologies.

In these systems, quantum circuits are executed by mapping logical gates to pulse-level controls on physically coupled qubit pairs. Accurate execution requires precise control of qubit–qubit interactions, activating couplings only when needed~\cite{beverland_assessing_2022, tomesh_quantum_2021, murali2020software}. However, as coupling density increases, frequency crowding makes it more difficult to isolate specific interactions, leading to degraded gate fidelity. At the same time, high connectivity is desirable to reduce \texttt{SWAP} overhead and support efficient circuit execution and error correction~\cite{murali_full-stack_2019}. Modular superconducting architectures partially address this tension by enabling selective, localized coupling within modules.

A central challenge in Noisy Intermediate-Scale Quantum (NISQ) architecture design is therefore balancing connectivity and fidelity (Fig.~\ref{fig:codesign-flow}). Sparse coupling reduces crosstalk and improves base gate fidelities, but increases routing overhead. Conversely, dense connectivity reduces \texttt{SWAP} insertion but introduces parasitic interactions that degrade performance. This tradeoff is further influenced by the coupler-to-qubit ratio. While cross-resonance-based architectures (e.g., IBM systems) employ constrained topologies such as heavy-hex lattices, other tunable-coupler designs, such as the Superconducting Nonlinear Asymmetric Inductive eLement (SNAIL), enable more flexible multi-qubit interactions through parametric coupling within a module, at the cost of residual spectator-induced errors~\cite{zhou_realizing_2023}. This flexibility significantly expands the architectural design space, provided that frequency crowding and spectator effects can be effectively managed.

In this paper, we present a physics-informed co-design framework that connects frequency allocation in tunable-coupler architectures with fidelity-aware circuit mapping. 
Using the SNAIL coupler as a core design element that determines modular size, we analyze the connectivity and gate fidelity trade-off by incorporating physical constraints---qubit frequency ranges, coupler bandwidths, and hybridization strengths---into an error budgeting framework to propose architectures meeting fidelity targets under realistic chip constraints~\cite{brink_device_2018, tripathi_operation_2019, ni_superconducting_2023, sete_error_2024}.  In particular, we explore how increasing the number of qubits per SNAIL affects the ability to selectively activate \iswap{} family gates. While prior work has optimized frequency allocation and fabrication-aware layout~\cite{li_towards_2019, ding_systematic_2020, smith_scaling_2022, morvan_optimizing_2022, osman_mitigation_2023, zhang_qplacer_2024, zhangEfficientFrequencyAllocation2024}, we focus on gate fidelity constraints induced by coupling all module qubits to a single SNAIL using its third-order mixing. 

Determining the impact of these architectural tradeoffs requires exploring the impact of NISQ quantum workloads for the different proposed architectures. However, to do this effectively requires a transpilation approach that can effectively trade off the gate error and circuit path lengths from \sw{} insertion due to topological flexibility. In particular, we propose a novel transpilation approach based on the SABRE algorithm~\cite{li2019tackling} extended with MIRAGE mirror gate insertion~\cite{mckinney2024mirage} to incorporate gate fidelity from the architecture module design choice in terms of the number of qubits and which gate pairs are supported in the module. We call this new transpilation pass FINESSE or Fidelity-INtegrated Equivalence-aware \sw{} SElection. 


FINESSE improves on SABRE/MIRAGE's randomized tie-breaking process for \sw{} placements which uses the downstream depth-optimizing approach when different candidates would locally see an identical path length.  Instead, FINESSE directly incorporates the impact of spectator aware noise to determine the minimum log of infidelity cost circuit in place of the solely circuit depth optimization approach. FINESSE's noise map also allows for noise models between devices to be included and to find the minimum noise mapping.

Our contributions in this paper include collaboratively designed improvements to the \textit{circuit mapping and topology} portion of the full quantum stack from determining the unique qubit frequency allocation to building a physics-informed transpilation noise model as illustrated in Figure~\ref{fig:codesign-flow} and to demonstrate the impact of this approach on the remainder of the design flow.  In particular our contributions include:

\begin{itemize}
\item We show that gate fidelity is \textbf{100$\times$ more sensitive to detuning for SNAIL-qubit conversion than for qubit-qubit conversion}, establishing fundamental limits on modular architecture design.
\item We solve a structured frequency allocation problem to determine the largest viable module size under fidelity constraints. We find \textbf{4 qubits per SNAIL} is the practical upper bound for preserving $>0.99$ fidelity.
\item We demonstrate a noise-aware transpilation approach from the solution to the structured frequency allocation problem to ensure an efficient mapping of circuits to the studied architecture, comparing to standard methods and find an \textbf{improvement in both accumulated log-infidelity cost and circuit depth}.
\item We conduct a study of different workloads onto architecture candidates to guide the design of module size and connectivity for system builds. 
As connectivity increases, noise degrades fidelity; consequently, a 4-qubit, 4-edge SNAIL fabric provides the best fidelity--connectivity tradeoff even when using noise-aware transpilation.
\end{itemize}

The remainder of this paper is organized as follows. Section~\ref{sec:background} provides background on two-qubit quantum gates, particularly in the context of the SNAIL coupler. SNAIL module infidelity is evaluated both intra- and inter-module in Section~\ref{sec:infidelity}. Section~\ref{sec:allocation} examines the impact of qubit and gate pair selection on gate fidelity. The FINESSE transpiler is presented in Section~\ref{sec:finesse}, and the evaluation of both the system architecture and transpiler is presented in Section~\ref{sec:results}. Finally, discussion and conclusions are provided in Section~\ref{sec:Conclusions}.
\section{Background}
\label{sec:background}
This section provides a brief overview of two-qubit gates such as \texttt{iSWAP}, introduces the SNAIL-based modular architecture used in our study, and defines the fidelity metrics used to evaluate architectural trade-offs.

\begin{figure}[tbp]
    \centering
    \includegraphics[width=\columnwidth]{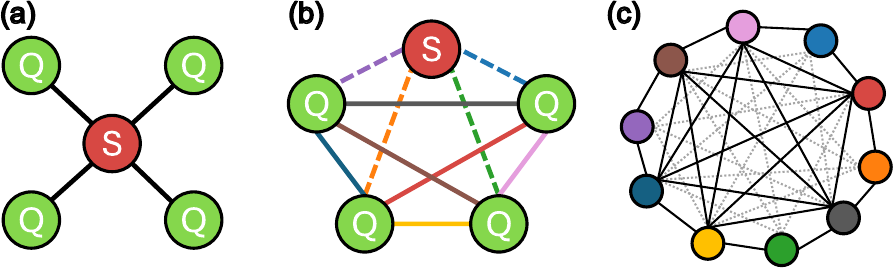}
    \caption{(a) SNAIL-qubit coupling graph (b) 2Q conversion connectivity graph, with unique colored edges for activated exchanges. (c) Compatibility graph, with nodes as interactions and edges as interferences.}
    \label{fig:colored-nodes}
\end{figure}

\begin{figure}[tbp]
    \includegraphics[width=\columnwidth]{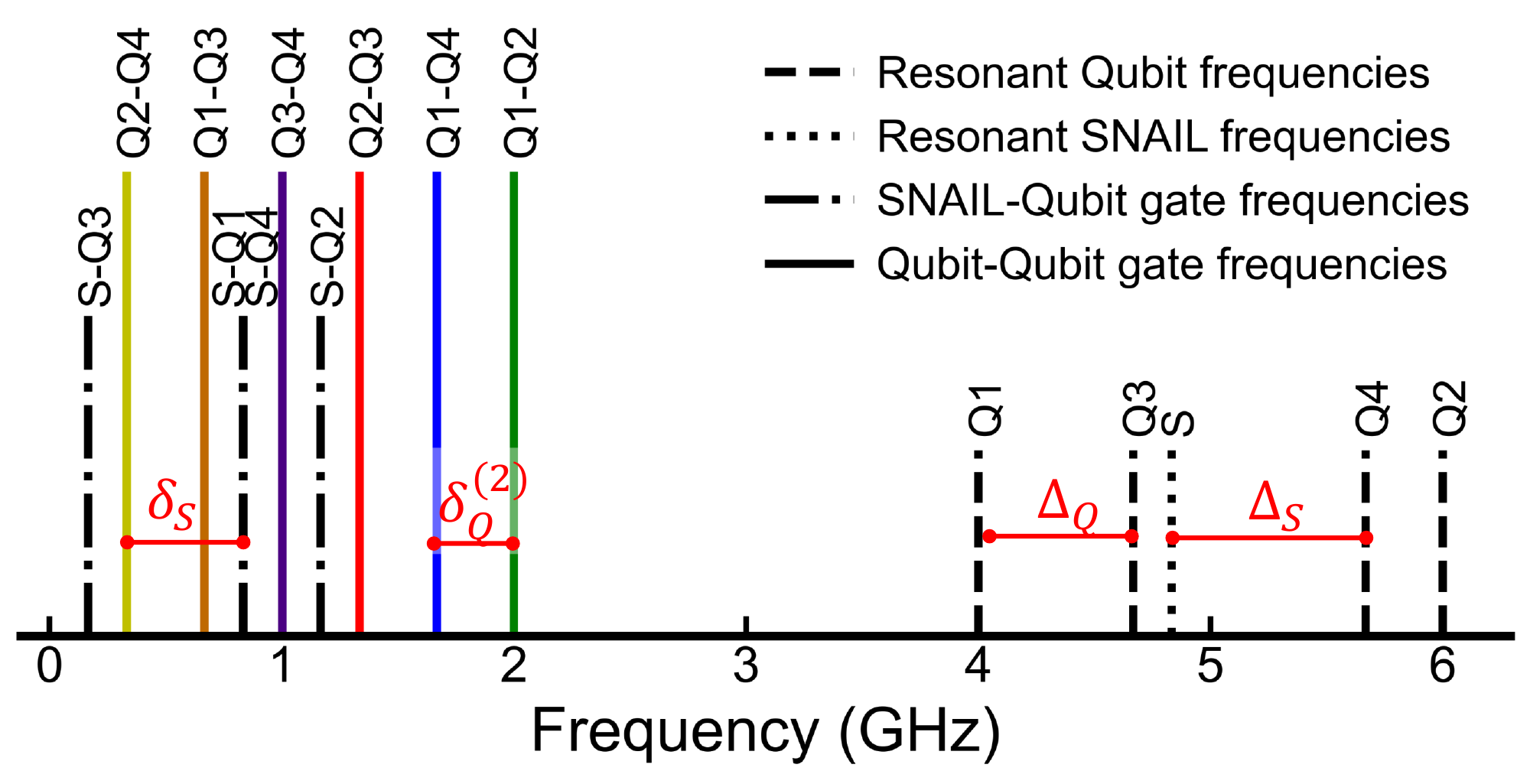}
    \caption{Spectral positioning of a SNAIL and 4 qubit bare modes and their interacting resonant frequencies.}
    \label{fig:spectral_crowding}
\end{figure}


Quantum computation is performed using unitary operations. Single-qubit gates are typically fast and low-error, while two-qubit gates introduce entanglement but are more error-prone. Though \cnot{} is common in algorithms, the \iswap{} gate is more naturally implemented in tunable coupler architectures enabling direct exchange of quantum states. Partial \iswap{}s can be realized by adjusting the pulse time applied to the coupler to realize a full family of \niswap{n} gates that realize the unitary:

\begin{equation}
    \label{iswap}
    \sqrt[n]{\texttt{iSWAP}} =
    \begin{bmatrix}
    1 & 0 & 0 & 0\\
    0 & \cos({\pi/2n}) & \mathit{i}\sin({\pi/2n}) & 0\\
    0 & \mathit{i}\sin({\pi/2n}) & \cos({\pi/2n}) & 0\\
    0 & 0 & 0 & 1
    \end{bmatrix}
\end{equation}

Gate dynamics are governed by the system Hamiltonian. Assuming time-independent evolution, the unitary is:
\begin{equation}
\label{eq:unitary}
U(t) = e^{-i \int_0^t H(t') dt' / \hbar}  = e^{-i \hat{H} t / \hbar}
\end{equation} 
Understanding the ideal gate unitary helps identify and mitigate deviations due to parasitic interactions or decoherence.


We implement two-qubit gates using a SNAIL as a tunable coupler~\cite{frattini17}. The SNAIL allows selective activation of specific interaction terms via targeted parametric pumping. A particular realized modular configuration of a SNAIL with transmon qubits is the ``Corral" module~\cite{zhou_realizing_2023, mckinney_co-designed_2023}, which contains four fixed-frequency transmons, each with a single Josephson junction (JJ) that provides dominant fourth-order nonlinearity~\cite{Koch07}. The SNAIL has an asymmetric loop of JJs, which when flux-biased to a critical value creates a device with strong third-order nonlinearity~\cite{frattini17}. Each qubit is capacitvely coupled to a readout resonator and controlled via dedicated drive lines.

Parametric drives at resonant conversion frequencies induce hybridization-mediated exchange between qubit pairs. Gate performance therefore depends on careful management of interaction frequencies to avoid unintended exchanges and spectator-induced errors (Figure~\ref{fig:colored-nodes}). Modules with varying numbers of qubits and SNAIL couplers can be realized through appropriate frequency tuning. As an example, Figure~\ref{fig:spectral_crowding} illustrates frequency allocation for the Corral architecture, where the conversion frequencies corresponding to driven interactions are shown as solid lines of differing colors. Ensuring correct gate operation requires sufficient frequency separation between SNAIL–qubit and conversion (qubit–qubit) frequencies, denoted by $\delta_S$, as well as separation among individual qubit–qubit interaction frequencies, denoted by $\delta_Q^{(2)}$.


Quantum circuits incur both coherent (unitary) and incoherent (dissipative) errors. Coherent errors stem from unwanted Hamiltonian terms—e.g., crosstalk and spectator interactions. Incoherent errors include energy relaxation ($T_1$) and dephasing ($T_\phi$), which irreversibly degrade state fidelity. 
In this work we construct a physics-informed fidelity model that captures both types of errors.
The average gate infidelity quantifies deviation from the target unitary~\cite{nielsen_simple_2002}:
\begin{equation} 
\label{eq:avg_fidelity}
\epsilon_{\text{avg}} = 1 - \frac{d + \text{Tr}[U V^\dagger]}{d +1},
\end{equation}
where $U$ is the ideal gate, $V$ is the noisy implementation, and $d$ is the Hilbert space dimension. Using the weak-coupling approximation where spectator-induced errors are treated as additive~\cite{murali_full-stack_2019}
, total coherent infidelity is:
\begin{equation}
\epsilon_{\text{coh}} \approx \sum_{i} \epsilon_\text{avg}^{(i)},
\end{equation}
where $\epsilon_\text{avg}^{(i)}$ denotes the infidelity from the $i$th spectator. We model incoherent decay using the standard exponential form derived from coherence times: $\epsilon_\text{inc} \approx 1 - e^{-t/\tau}$. Treating coherent and incoherent processes as approximately independent error channels, the combined infidelity is given by:
\begin{equation}
\label{eq:total_fidelity}
\epsilon_{\text{gate}} \approx 1 - (1-\epsilon_\text{inc}) \times (1 - \epsilon_{\text{coh}}).
\end{equation}
This decomposition isolates unitary and non-unitary effects: spectator-induced infidelities add, while decoherence contributes a multiplicative penalty. This model underlies our cost function for frequency allocation~\cite{hopf2025improving, gokhale2024faster, schmid2024computational} and is discussed in Section~\ref{sec:allocation}. Background on SABRE-family transpilation strategies, used to evaluate the architectural choices emerging from the frequency allocation study, is provided in Section~\ref{sec:finesse}.
\section{Infidelity in a Module}
\label{sec:infidelity}
Characterization of how different types of infidelity contribute to errors within a module requires quantification of the dependence of gate infidelity on frequency detuning for relevant interactions. The dominant error sources arise from spectator terms, which accumulate as coherent infidelity. In addition, subharmonic constraints of the SNAIL coupler limit achievable gate speeds, contributing to incoherent infidelity. Together, these effects establish the frequency separation constraints used in subsequent chip-level frequency allocation.


\subsection{Hamiltonian Expansion}
To derive the two-qubit basis gates, the static system Hamiltonian forms the basis on which standard transformations can push into the interaction frame, allowing each term to acquire a pump-frequency-dependent coefficient.

The SNAIL-Corral architecture implements an \iswap{} 
instruction set~\cite{chen_one_2023} using photon-conversion interactions. The full Hamiltonian is decomposed as $H = H_{0L} + H_{NL} + H_c$, with $H_{0L}$ capturing the bare mode frequencies:
\begin{equation}
H_{0L} = \omega_s s^\dagger s + \sum_{i} \omega_i q_i^\dagger q_i    
\end{equation}
where $\omega_s$ and $\omega_i$ denote the SNAIL and qubit frequencies. The nonlinear terms are:
\begin{equation}
H_{NL} = g_3 (s^\dagger + s)^3 + \sum_{i} \frac{\alpha_i}{12}(q_i^\dagger + q_i)^4
\end{equation}
with $g_3$ the third-order SNAIL nonlinearity and $\alpha_i$ the transmon anharmonicity. Coupling between SNAIL and qubits is defined by:
\begin{equation}
H_c = \sum_q g_{sq} (s^\dagger q + s q^\dagger),
\end{equation}
where $g_{sq}$ is the coupling strength.

The interaction frame is useful for describing an effective Hamiltonian where each interaction term is modulated by a rate (the strength of the pump) and a complex phase factor (how far off-resonance the interaction is from the pump's frequency). Transforming to the interaction frame via the Bogoliubov diagonalization and displacement transform~\cite{zhou_superconducting_2023,zhou_realizing_2023, xia2023fast}, the Hamiltonian becomes:
\begin{multline}
\label{eq:effective-hamiltonian}
\tilde{H}_I = g_3 \left( s e^{-i \tilde{\omega}_s t} + \eta e^{-i \omega_p t} + \sum_{i} \lambda_{si} q_i e^{-i \tilde{\omega}_{q_i} t} + \text{h.c.} \right)^3 \\
\quad + \sum_{i} \frac{\alpha_i}{12} \left( q_i e^{-i \tilde{\omega}_{q_i} t} - \lambda_{si}(s e^{-i \tilde{\omega}_s t} + \eta e^{-i \omega_p t}) + \text{h.c.} \right)^4\end{multline}

Here, $\tilde{\omega}$ are dressed mode frequencies, and $\lambda_{si} = g_{sq}/\Delta$ quantifies hybridization between the SNAIL and each qubit. The pump strength $\eta = \sqrt{n_s}$ reflects the coherent pump amplitude in the SNAIL mode. Terms are resonant when $\tilde{\omega}_x \approx \omega_p$, and vanish when far-detuned. The Rotating-Wave Approximation (RWA) retains only those terms near resonance, this \textit{assumes that the unwanted terms are actually far enough off-resonance} and thus could be considered sufficiently small~\cite{zhou_superconducting_2023,zhou_realizing_2023, xia2023fast,barajas2025quantum}.

We use OpenFermion and SymPy to symbolically expand the cubic and quartic Hamiltonian polynomials (Eq.~\ref{eq:effective-hamiltonian}), in addition to normal-ordering and expression simplification. We retain dominant terms: two-qubit \iswap{}
s, SNAIL-qubit conversions, and subharmonic single-mode drives (Table~\ref{tab:order_magnitude_terms}). Example omitted terms 
are gain-process terms, conjugate operators created by frequency additions rather than differences, because they are far from the pump.

\begin{table}[t]
\centering
\caption{Order of magnitude analysis for driven, intra-module, and inter-module spectator terms sorted by normalized prefactor. Here, $q_a$ and $q_b$ are generic qubits in the driven module; $s$ denotes the SNAIL; and $q_c$, $q_d$, $s_n$ denote neighboring module elements.}
\resizebox{\columnwidth}{!}{
\begin{tabular}{|c|c|c|c|}
\hline
\multicolumn{4}{|c|}{\textbf{Driven Term}} \\ \hline
Term & Coefficient & $\omega_p =$ & Normalized Prefactor \\ \hline
$(q_a^\dagger q_b + q_a q_b^\dagger)$ & $6 |\eta| \lambda^2 g_3$ & $|\omega_{q_b} - \omega_{q_a}|$ & 1.0 \\ \hline
\multicolumn{4}{|c|}{\textbf{Intra-Module Spectator Terms}} \\ \hline
$(s^\dagger + s)$ & $3 |\eta|^2 g_3$ & $\omega_s / 2$ & 100.0 \\ \hline
$(s^\dagger q_a + s q_a^\dagger)$ & $6 |\eta| \lambda g_3$ & $|\omega_s - \omega_{q_a}|$ & 10.0 \\ \hline
$(q_a^\dagger + q_a)$ & $3 |\eta|^2 \lambda g_3$ & $\omega_{q_a} / 2$ & 10.0 \\ \hline
$(s^\dagger q_a + s q_a^\dagger)$ & $\alpha |\eta|^2 \lambda^3$ & $|\omega_s - \omega_{q_a}| / 2$ & 0.067 \\ \hline
$(q_a^\dagger + q_a)$ & $\alpha |\eta|^3 \lambda^3/3$ & $\omega_{q_a} / 3$ & 0.044 \\ \hline
$(s^\dagger + s)$ & $N_q \alpha |\eta|^3 \lambda^4/3$ & $\omega_s / 3$ & 0.018 \\ \hline
\multicolumn{4}{|c|}{\textbf{Inter-Module Spectator Terms}} \\ \hline
$(s_n^\dagger + s_n)$ & $3 |\eta|^2 \lambda^2 g_3$ & $\omega_{s_n} / 2$ & 1.0 \\ \hline
$(s^\dagger q_c + s q_c^\dagger)$ & $6 |\eta| \lambda^3 g_3$ & $|\omega_s - \omega_{q_c}|$ & 0.1 \\ \hline
$(q_c^\dagger + q_c)$ & $3 |\eta|^2 \lambda^3 g_3$ & $\omega_{q_c} / 2$ & 0.1 \\ \hline
$(q_a^\dagger q_c + q_a q_c^\dagger)$ & $6 |\eta| \lambda^4 g_3$ & $|\omega_{q_c} - \omega_{q_a}|$ & 0.01 \\ \hline
$(s_n^\dagger q_a + s_n q_a^\dagger)$ & $6 |\eta| \lambda^5 g_3$ & $|\omega_{s_n} - \omega_{q_a}|$ & 0.001 \\ \hline
$(q_c^\dagger q_d + q_c q_d^\dagger)$ & $6 |\eta| \lambda^6 g_3$ & $|\omega_{q_d} - \omega_{q_c}|$ & $0.0001$ \\ \hline
\end{tabular}
}
\label{tab:order_magnitude_terms}
\end{table}

To isolate a desired two-qubit gate, we set the pump on-resonance, $\omega_p = \tilde{\omega}{q_2} - \tilde{\omega}{q_1}$, selecting the term:
\begin{equation}
H_{\text{target}} = 6 |\eta| g_3 \lambda^2 (q_1 q_2^\dagger + q_1^\dagger q_2)
\end{equation}
The required pulse duration $t_f$ follows from Eq.~\ref{eq:unitary}, yielding:
\begin{equation}
\label{eq:iswap-eta}
\frac{\pi}{2n} = 6 t_f |\eta| g_3 \lambda^2 \quad \text{(}\sqrt[n]{\text{iSWAP}}\text{)}
\end{equation}

\begin{figure}[tbp] 
    \centering 
    \includegraphics[width=.6\columnwidth]{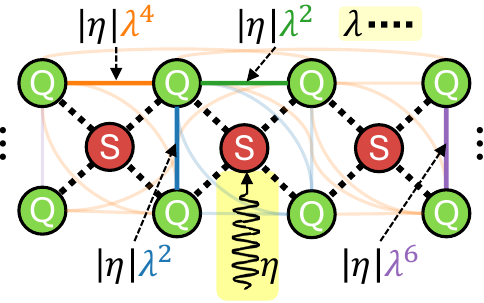} 
    \caption{In this diagram, the central SNAIL is driven with the target interaction denoted by a green edge. Spectator terms diminish based on how many orders of hybridization they are removed from the driven SNAIL. (Blue) Both qubits coupled to the driven SNAIL (in the same module); (Orange) One qubit directly coupled to the driven SNAIL; (Purple) Neither qubit directly coupled to the driven SNAIL.} 
    \label{fig:corral}
\end{figure}

Spectator interactions result from residual hybridization between qubits and the SNAIL. We distinguish intra-modular spectators---where the qubit is directly coupled to the driven SNAIL---from inter-modular ones, which involve indirect pathways through neighboring SNAILs (Figure~\ref{fig:corral}). The strength and detuning of these interactions determine their contribution to infidelity, as detailed in Table~\ref{tab:order_magnitude_terms}.

\subsection{Spectator Infidelities}
\label{sec:simulation-constraints}
To establish frequency separation constraints, the impact of detuning on gate fidelity is evaluated. Coherent errors arise from unintended spectator interactions, while incoherent errors are associated with limitations on SNAIL coupler speed.

\subsubsection{Coherent loss from spectators}
\begin{figure}
    \centering
    \includegraphics[width=\columnwidth]{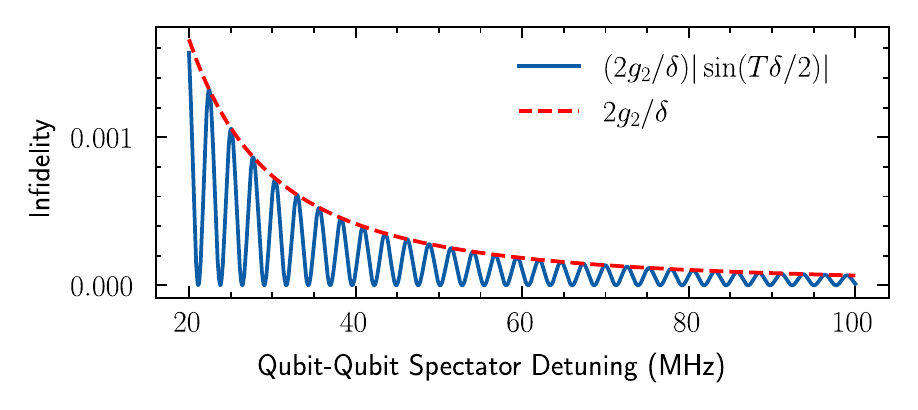}
    \caption{Infidelity versus detuning for spectator amplitude with Rabi oscillations compared with spectator amplitude using Rabi magnitude bound.}
    \label{fig:detuning_rabi}
\end{figure}

To quantify spectator-induced infidelity, each interaction term is isolated and its impact is evaluated using Eq.~\ref{eq:avg_fidelity}. A representative Hamiltonian with one target and one detuned spectator gate is:
\begin{equation} 
H(t) = 6 |\eta| g_3 \lambda^2 \left( q_1^\dagger q_2 + e^{-it \delta_Q} q_3^\dagger q_4 + \text{h.c.} \right)
\end{equation}
Here, $\delta_Q$ denotes the detuning between the spectator term and the target gate.

\begin{figure}[t]
    \centering
    \includegraphics[width=.8\columnwidth]{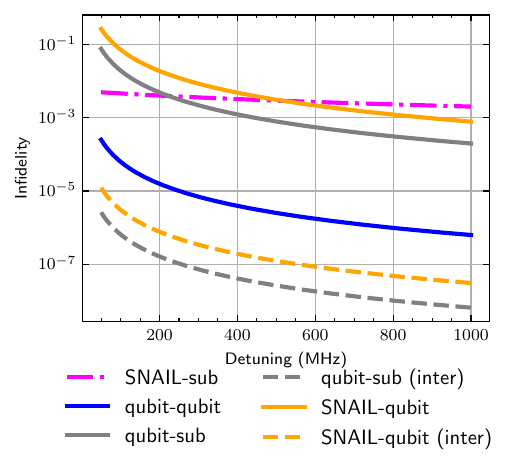}
    \caption{Infidelity scaling versus detuning for different spectator types.}
    \label{fig:fidelity_vs_terms}
\end{figure}


In certain cases, the target gate may coincide with the anti-node of the spectator oscillation (Figure~\ref{fig:detuning_rabi}), partially mitigating its effect. However, in realistic devices where interaction rates and gate durations are not precisely synchronized due to fabrication variability.
Instead, we apply a conservative amplitude bound:
\begin{equation}
\left| \int_0^{T} e^{-i t \delta} dt \right| = 2 | \sin (T \delta /2) |  / \delta \leq 2/\delta
\end{equation}
This avoids dependence on precise Rabi synchronization and simplifies the simulation by removing time dependence from the interaction term. The resulting unitary becomes:
\begin{equation}
\label{eq:spectator}
U(t) = e^{-i \left (g_1 t (q_1^\dagger q_2) + 2 g_2  (q_3^\dagger q_4)  / \delta  + h.c. \right)}
\end{equation}
with $g_1$ and $g_2$ determined by the target and spectator coefficients in Table~\ref{tab:order_magnitude_terms}. While the example above uses a spectator qubit-qubit term $q_3^\dagger q_4 + q_3 q_4^\dagger$, the same procedure can be applied to other off-resonant interactions in a similar fashion. The corresponding infidelity is then computed across a range of $\delta_Q$ values and fit to an empirical model:
\begin{equation}
\label{eq:coh_cost}
\epsilon_\text{coh}(\delta) := \frac{2 x_0}{(x_1 + \delta)^2}  
\end{equation}  
where $x_0,x_1$ serve as fit parameters, as shown in Figure~~\ref{fig:fidelity_vs_terms}.

\subsubsection{Incoherent loss from speed limits}
\label{sec:snail-death}

\begin{figure}[t]
    \centering
    \includegraphics[width=\columnwidth]{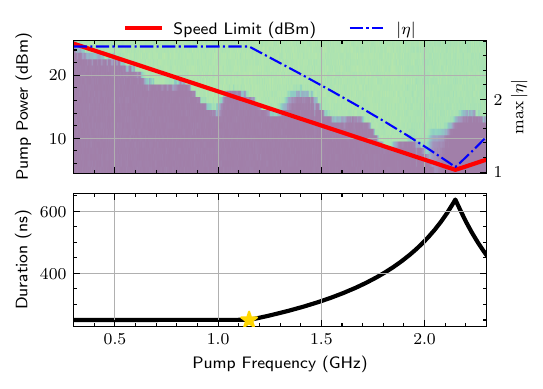}
    \caption{(a) Maximum pump power vs. pump frequency. Overlaid is the maximum achievable $|\eta|$, fitted under the assumption that a viable gate duration exists at the star marker. (b) Gate duration vs. pump frequency, using the star marker to establish a scaling factor between pump power and $|\eta|$.}
    \label{fig:snail_limits}
\end{figure}

We next consider incoherent loss from pump power constraints. Figure~\ref{fig:snail_limits} shows experimental SNAIL breakdown data~\cite{zhou_superconducting_2023, mckinney2023parallel}, where exceeding a pump threshold causes loss of parametric behavior. Although the detailed physics are complex, we fit this data to approximate the maximal achievable $\eta$ as a function of pump frequency.

Driving the SNAIL resonantly enables gate operations by populating it with photons. However, excessive pump power causes breakdown, particularly in subharmonic modes that are $\sim$100$\times$ stronger than typical interaction terms (Table~\ref{tab:order_magnitude_terms}). While these do not directly introduce unitary error, they destabilize gate execution and affect layout density and parallelization~\cite{dumas2024unified}.

From Eq.~\ref{eq:iswap-eta}, gate duration $t_f$ is determined by $\eta$, which itself depends on $\omega_p$ and $\omega_s$:
\begin{equation}
|\eta| = \frac{\epsilon \omega_s}{\omega_p^2 - \omega_s^2},
\end{equation}
Maintaining fixed gate time at low $\omega_p$ requires increasing drive amplitude $\varepsilon$, trading off gate speed against stability.

To model this tradeoff, we assume a linear relationship between dBm and $\varepsilon$, justified by attenuation being roughly linear across the drive chain. This allows us to define a breakdown threshold where increasing $\varepsilon$ further would destabilize the coupler. Above this threshold, additional speedup is infeasible. The critical photon number~\cite{frattini2021three} limits how strongly the SNAIL can be driven. Beyond this, nonlinear effects and chaotic behavior disrupt gate performance~\cite{xia2023fast}. Other second-order effects---such as pump-induced frequency shifts and higher-order Kerr terms---further limit fidelity but are not modeled explicitly here~\cite{Raman2018}.


Assuming a viable gate exists at the star marker in Figure~\ref{fig:snail_limits} (250 
ns at 1 GHz detuned from $\omega_S/2$), we fit the maximum usable $\eta$ curve and compute $t_f$. Using Eq.~\ref{eq:total_fidelity}, we convert gate duration into decoherence-induced loss and fit the result with the 
following empirical approximation:
\begin{equation}
\label{eq:incoh_cost}
\epsilon_\text{inc}(\delta) := \frac{x_0}{(x_1 + \delta)}
\end{equation}
again using $x_0,x_1$ as fit parameters, which is plotted in Figure~\ref{fig:fidelity_vs_terms}.

\subsection{Design Tradeoffs}
\label{sec:tradeoffs}
\begin{figure}[tbp]
    \centering
    \includegraphics[width=\columnwidth]{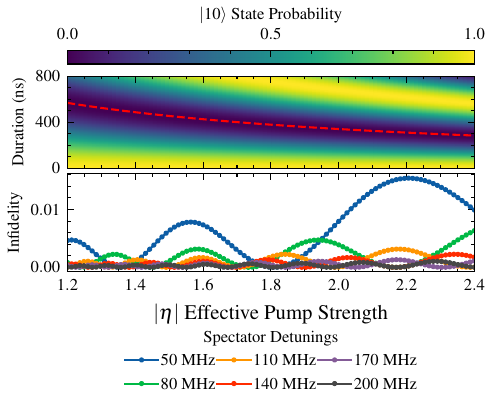}
    \caption{(a) Population exchange between $|01\rangle$ and $|10\rangle$ as a function of time and pump strength. The dashed red line marks a full \iswap{}. (b) Coherent infidelity from a qubit-qubit spectator vs. effective pump strength at various $\delta_Q$.}
    \label{fig:infidelity_vs_eta}
\end{figure}

Optimizing gate fidelity requires balancing gate speed against spectator interactions. Stronger pumps shorten gate time but amplify unwanted interactions (Figure~\ref{fig:infidelity_vs_eta}). From Eq.~\ref{eq:spectator}, achieving $g_1 t_f \gg 2g_2 / \delta_Q$ ensures the target gate dominates. Increasing $\delta_Q$ suppresses spectators but lengthens $t_f$, increasing $T_1$-induced loss.

\begin{figure}[tbp]
    \centering
    \includegraphics[width=\columnwidth]{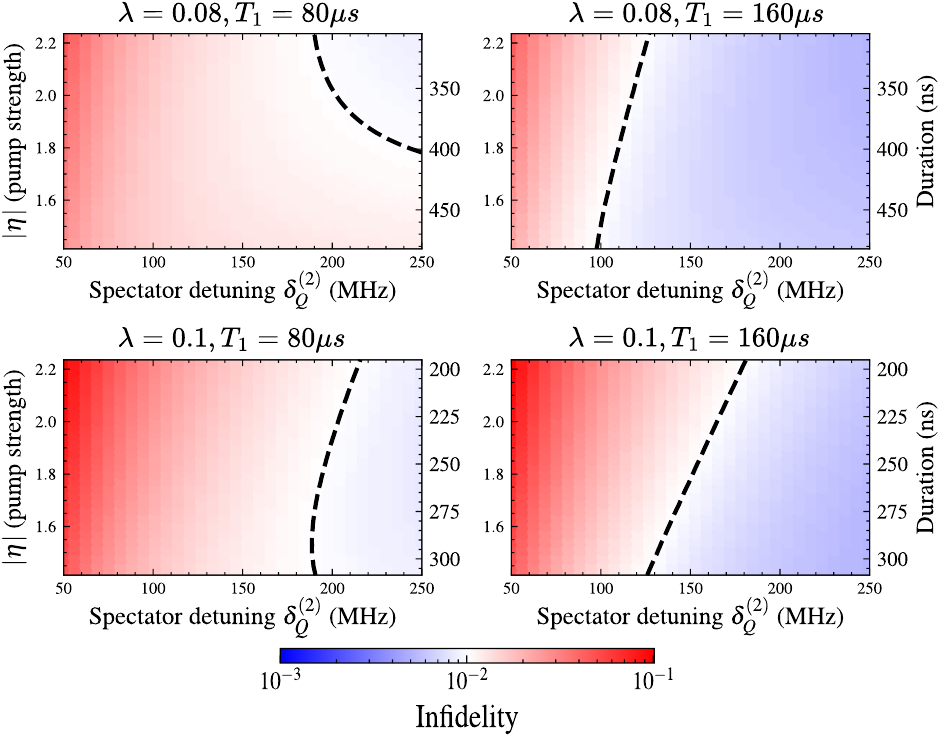}
    \caption{Comparison of \iswap{} fidelity threshold boundaries for different values of $\lambda$ and $T_1$.}
    \label{fig:combined-results}
\end{figure}

Using QuTiP's Lindblad solver to evaluate both error types simulates noisy evolution with amplitude damping for $T_1$. The resulting operator $U(t_f)$ defines the fidelity threshold region in Figure~\ref{fig:combined-results}, plotted over $\lambda$ and $T_1$. High-fidelity gates ($F \geq 0.99$) appear after around 150–200 MHz detuning. Weaker couplings shift the threshold left, into $T_1$-dominated loss; stronger pumps shift it upward, requiring more detuning to suppress spectators. This reveals an optimal operating region where both error sources are balanced---supporting the bounds in Figure~\ref{fig:snail_limits}, which limit $\eta$ accordingly.

\begin{sloppypar}
We assume uniform SNAIL-qubit coupling ($\lambda_i \equiv \lambda_j$), though fabrication variability causes nonuniform hybridization. This affects both gate duration and spectator interference (Figure~\ref{fig:detuning_rabi}). Higher-order nonlinearities were not modeled in the transmon potential, which can cause AC Stark shifts and access higher excitation levels. Direct qubit-qubit and SNAIL-SNAIL couplings ($\lambda_{qq} = \lambda_{ss} = 0$) were also ignored, though these may be non-negligible when modules are densely packed. In particular, couplers sharing qubits can introduce residual cross-coupling due to their physical proximity, complicating interaction isolation.
\end{sloppypar}

\section{Qubit Frequency Allocation}
\label{sec:allocation}

\begin{figure}[tbp]
    \centering
    \includegraphics[width=\columnwidth]{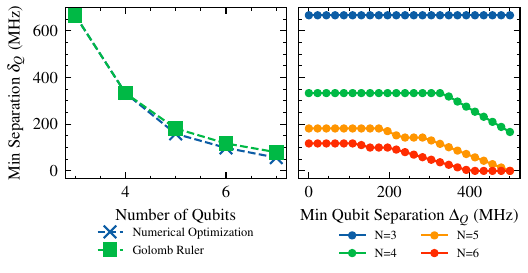}
    \caption{(Left) Minimum Difference Separation vs. Number of Qubits. Numerical optimization scales with the analytical Golomb Ruler but trails the optimal solution due to slight instability in convergence. (Right) Separation results with additional constraint of minimum allowed spacing between bare qubit frequencies.}
    \label{fig:interaction_separation_combined}
\end{figure}

The infidelity models from Section~\ref{sec:simulation-constraints} can be used to construct frequency allocations that maximize gate fidelity. This allocation task can be formulated as a variant of the classical Frequency Assignment Problem (FAP) and solved using numerical optimization. The resulting solutions enable evaluation of trade-offs that arise in multi-module layouts.

FAP is a well-known NP-complete problem in which discrete frequency channels are assigned to transmitters while avoiding interference~\cite{park_application_1996, waters2005graph, orden_spectrum_2018}. It is typically cast as a graph coloring problem: nodes are transmitters, colors are frequencies, and edges encode minimum required spacing:
\begin{equation} 
|f(i) - f(j)| \geq q_{ij} \quad \forall (i, j), i \neq j 
\end{equation}

While the classical objective is often to minimize total bandwidth, alternate formulations maximize the minimum pairwise spacing:
\begin{equation} 
\max \min_{i, j \in X, i \neq j} |f(i) - f(j)| 
\end{equation}

Quantum frequency allocation differs fundamentally from classical FAP in that physical constraints act on differences between frequencies, not their absolute values. Two-qubit gates are mediated by frequency detunings, and nonlinearities like subharmonic and hybridization effects impose coupled constraints across all interactions. These dependencies create a rugged optimization landscape that cannot be decomposed into independent frequency assignments.


To quantify achievable interaction separation, we establish a theoretical baseline using a Golomb ruler as an optimal bound. A Golomb ruler is a sequence of marks (interpreted here as frequencies) where all pairwise differences are unique, minimizing spectral crowding~\cite{memarsadeghi2016nasa}. One construction is given by:
\begin{equation}
f_k = f_\text{min} + c (2pk + (k^2 \bmod p)), \quad k = 0, 1, \dots, p-1
\end{equation}
where $p$ is the number of qubits and $c$ a scaling factor. Though suspected NP-hard~\cite{meyer2009complexity}, optimal sequences are known for small $p$, providing a benchmark for interaction separation.

However, the Golomb sequence requires marks to become increasingly close together, focusing only on relative differences while ignoring absolute placement constraints. In quantum systems, overly-close qubit frequencies create new challenges; for example, if qubits are too close in frequency, single-qubit gates become difficult to address, and higher-order state transitions introduce additional interference. To account for this, we introduce a constraint $\Delta_Q$ on the minimum allowed spacing between bare qubit frequencies.

\begin{algorithm}[htb]
\caption{Minimize Spectator Infidelity}
\label{alg:numerical_optimizer}
\begin{algorithmic}[1]
\Require Module graph, frequency bounds $\omega_Q, \omega_S$,$\epsilon_{\text{coh}}, \epsilon_{\text{inc}}$, worst-gate exclusion $k$
\Ensure Optimized frequency assignment minimizing total infidelity

\State Initialize random frequencies $\{ \omega_{q_i} \}, \omega_S$
\Repeat
    \State Collect all spectator frequencies $\mathcal{S} = \{ \omega_{\text{spec}} \}$ for all SNAIL-qubit, qubit-qubit pairs and qubit subharmonics
    \For{each gate $(q_i, q_j)$ in module}
        \State $\epsilon_{\text{coh}}(q_i, q_j) \gets 0$
        \For{each spectator $\omega_{\text{spec}} \in \mathcal{S}$}
            \State $\epsilon_{\text{coh}}(q_i, q_j) \mathrel{+}= \epsilon_{\text{coh}}^{\text{spec}}(|\omega_{q_i} - \omega_{\text{spec}}|)$
        \EndFor
        \State $\epsilon_{\text{inc}}(q_i) \gets \epsilon_{\text{inc}}(|\omega_{q_i} - \omega_S / 2|)$
        \State $\epsilon_{\text{gate}}(q_i, q_j) \gets 1 - (1 - \epsilon_{\text{coh}}(q_i, q_j))(1 - \epsilon_{\text{inc}}(q_i))$
    \EndFor
    \For{each qubit}
        \State Apply penalty if $\min |\omega_{q_i} - \omega_{q_j}| < \Delta_Q$
    \EndFor
    \State Sort $\{ \epsilon_{\text{gate}} \}$ in descending order and drop worst $k$ gates
    \State $\mathcal{L} \gets \sum_{\text{gates remaining}} \epsilon_{\text{gate}}$
    \State Update $\{ \omega_{q_i} \}, \omega_S$ using Nelder-Mead to minimize $\mathcal{L}$
\Until{convergence or max iterations reached}
\State \Return Optimized $\{ \omega_{q_i} \}, \omega_S$
\end{algorithmic}
\end{algorithm}

Figure~\ref{fig:interaction_separation_combined} compares the optimized frequency allocations to the Golomb bound. For small modules, the numerical solutions closely match the ideal separation. However, when $\Delta_Q$ constraints are enforced, the achievable spacing begins to deviate from the Golomb bound. For four qubits, the optimizer maintains separations exceeding \SI{300}{\mega \hertz}, while for five qubits the minimum pairwise spacing drops below \SI{200}{\mega \hertz}. This trend highlights the challenge of simultaneously preserving single-qubit isolation and sufficient detuning for high-fidelity two-qubit gates.

Unlike the linear FAP, the proposed cost function is nonlinear and incorporates absolute-value terms, detuning-dependent infidelity models, and coupled penalty terms. A numerical optimization procedure (Algorithm~\ref{alg:numerical_optimizer}) is used to assign qubit and coupler frequencies that minimize total two-qubit gate infidelity, accounting for both coherent (spectator-induced) and incoherent (lifetime-related) error contributions. Related approaches have explored alternative formulations, including graph neural network-based methods~\cite{ai2024graph} and perturbative correction techniques~\cite{mammola2025optimal}.

The scaling of spectator effects is derived from order-of-magnitude estimates in Table~\ref{tab:order_magnitude_terms}, which categorize the impact of different interaction types. Figure~\ref{fig:fidelity_vs_terms} translates these estimates into infidelity scaling, allowing the optimizer to focus on mitigating the most significant error sources. The largest contributors are coupler-qubit and qubit subharmonic, followed by qubit-qubit interactions. The full optimization process minimizes total infidelity for each gate:
\begin{equation} 
\epsilon_{\text{gate}}(q_i, q_j) = 1 - (1 - \epsilon_{\text{coh}}(q_i, q_j))(1 - \epsilon_{\text{inc}}(q_i)). 
\end{equation}
under the following constraints:
\begin{align}
&3.3 \text{ GHz} \leq \omega_Q \leq 5.7 \text{ GHz}\\
&4.2 \text{ GHz} \leq \omega_S \leq 4.7 \text{ GHz}\\
&|\omega_{q_i} - \omega_{q_j}| \geq \Delta_Q = 200\text{ MHz}.
\end{align}
which reflect operating regimes observed in our experimental studies of SNAIL-based systems.

\begin{figure}[tbp]
    \centering 
    \includegraphics[width=\columnwidth]{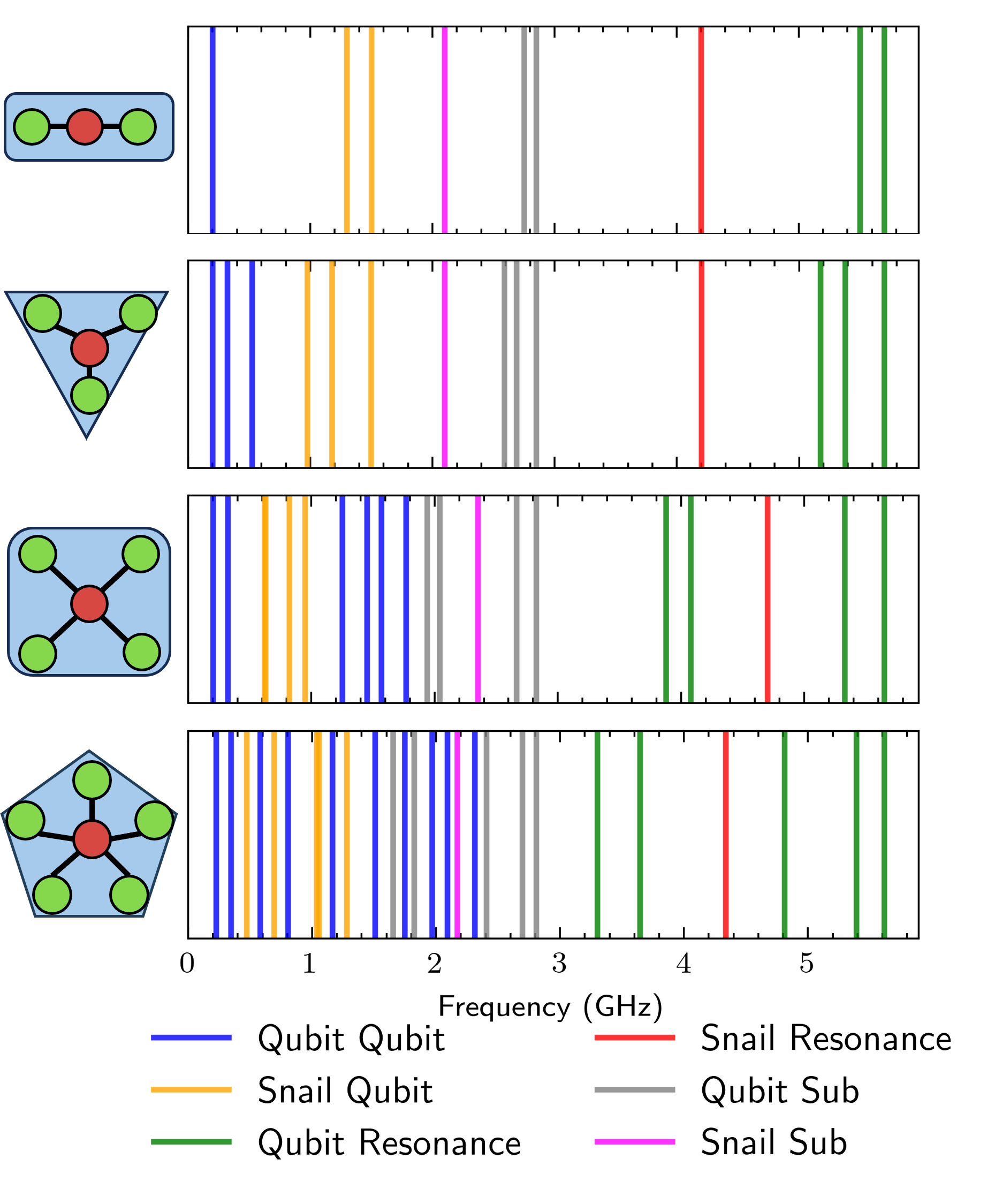}
    \caption{Optimized Frequency Stack: Frequencies of qubit and SNAIL resonances alongside interaction terms, grouped by module size $N = 2, 3, 4, 5$ (from top to bottom).}
    \label{fig:frequency_stack}
\end{figure}

\begin{figure}[tbp]
    \centering
    \includegraphics[width=\columnwidth]{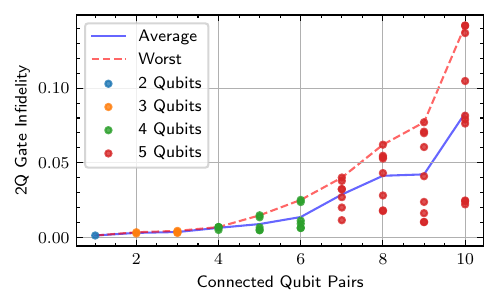}
    \caption{Average two-qubit gate infidelities across module sizes with and without selective edge removal.}
    \label{fig:base-fidelities}
\end{figure}

The optimization proceeds via Nelder--Mead, with heavy penalties imposed for violations of $\Delta_Q$. Figure~\ref{fig:frequency_stack} shows the optimized frequency placements, while Figure~\ref{fig:base-fidelities} reports gate fidelities across module sizes. The geometric mean gate fidelity for each module size is:
\begin{center}
$N\in\{2,3,4,5\}$: $1-\epsilon_{\text{gate}}\approx 0.996,\,0.994,\,0.991,\,0.940$
\end{center}
As expected, fidelity degrades with increasing module size due to tighter spectral constraints.

\section{FINESSE: Fidelity-Aware Routing}
\label{sec:finesse}
A central challenge in transpiling quantum circuits for NISQ systems is the selection of an objective function that accurately reflects resulting circuit fidelity under non-uniform gate performance. In current devices, gate fidelity varies due to multiple noise sources, including frequency-dependent spectator interactions and calibration drift. Effective transpilation frameworks, such as SABRE~\cite{li2019tackling}, typically optimize proxy metrics like circuit depth or \sw{} count to simplify the optimization process. These approaches implicitly assume uniform error contributions across gates. As demonstrated in Section~\ref{sec:allocation}, this assumption does not hold in current devices, where two-qubit gate infidelities can vary across physical qubit pairs, even when managed by the same coupler.

Transpilation can instead incorporate noise-aware cost models that more directly reflect circuit fidelity by jointly accounting for gate error, circuit depth, and heterogeneity in qubit-pair fidelities. For example, Figure~\ref{fig:base-fidelities} illustrates variation in gate fidelity across qubit pairs. A fidelity-aware transpiler can preferentially avoid low-fidelity interactions when possible; however, constraints imposed by circuit structure and limited connectivity introduce trade-offs between path length and access to high-fidelity gates. For instance, longer circuit paths may require higher-fidelity gates due to the accumulation of dependent operations, while shorter paths may tolerate lower-fidelity qubit pairs. Effectively navigating these trade-offs requires a cost function that captures gate-dependent error while accounting for circuit depth.

\subsection{Existing SABRE-based Heuristics}
\label{sec:SABRE}
SABRE~\cite{li2019tackling} represents a circuit as a directed acyclic graph (DAG) and maintains a \emph{front layer}~$F$ of ready gates and an \emph{extended set}~$E$ of the next $|E|$ successors (default $|E|=20$). When no front-layer gate is directly routable, SABRE evaluates candidate \sw{}s on coupling-map edges involving front-layer qubits using the heuristic:
\begin{equation}
\label{eq:sabre_H}
\!\!H = \max(\delta_{p_0}, \delta_{p_1})\!\left(\!\tfrac{1}{|F|}\!\sum_{g\in F} D[\pi(g)] \!+\! W\tfrac{1}{|E|}\!\sum_{g\in E} D[\pi(g)]\!\right)
\end{equation}
where $D[i,j]$ is the shortest-path distance between physical qubits $i$ and $j$, $\pi$ maps logical to physical qubits, $W=0.5$ weights the lookahead term, and $\delta_p$ penalizes recently swapped qubits to promote parallelism. The \sw{} minimizing $H$ is selected, with ties broken randomly.

SABRE begins from a random initial layout. A forward routing pass produces a complete mapping of logical outputs to physical qubits, followed by a reverse pass on the inverted circuit to refine the initial placement. This bidirectional process is iterated to improve the layout~\cite{li2019tackling}. LightSABRE~\cite{Zou2024} extends SABRE with \emph{relative scoring}, evaluating candidate \sw{}s by their change in the heuristic rather than recomputing the full cost, and introduces a \emph{release valve} that revisits long \sw{} chains ($\geq 10$) to ensure progress.

MIRAGE~\cite{mckinney2024mirage} extends SABRE by exploiting \emph{mirror gates} to reduce \sw{} count and improve decomposition efficiency. A mirror gate replaces $U$ with $U'=\mathrm{SWAP}\cdot U$, implicitly permuting outputs. During routing, when a \sw{} on edge $(p_0,p_1)$ is followed by a front-layer gate $U$ on the same edge, the two can be fused into a single mirror operation $U'$, simultaneously routing and executing the gate.  Under KAK decomposition, a two-qubit unitary $U$ requires $k_U$ applications of the native basis gate; the mirror $U'$ may admit a different cost $k_{U'}$. MIRAGE accepts the substitution when $k_{U'} \leq k_U$, reducing decomposition cost without increasing the \sw{} count.

To realize MIRAGE, the algorithm introduces an \emph{intermediate layer} between the execution and mapped layers of SABRE’s workflow, where each routable two-qubit gate is evaluated for mirror substitution prior to commitment. This decision is controlled by \emph{aggression levels}: (0) always rejecting mirrors; (1) accepting only if depth is reduced; (2) accepting if depth does not increase (default); and (3) always accepting mirrors. 

\subsection{Fidelity-Aware \sw{} Selection}
\label{sec:motivation}
\label{sec:FINESSE}

As previously discussed, $H$ in SABRE uses a proxy for circuit fidelity based on \sw{} insertion, commonly referred to as hop-count distance. This metric does not account for hardware-dependent gate fidelity, even though Figure~\ref{fig:base-fidelities} shows that coupling variation significantly impacts gate performance under different mappings. To address this limitation, we introduce \textbf{FASST} (Fidelity-Aware \sw{} Selection Transpilation), which modifies $H$ to become $H'$ by replacing hop-count distances with fidelity-weighted distances. 

Building on this, we introduce \textbf{FINESSE}, which reintegrates mirror-gate selection from MIRAGE into the fidelity-aware framework of FASST. By incorporating fidelity into both the lookahead heuristic and mirror acceptance, FINESSE makes \sw{} selection and gate equivalence decisions jointly fidelity-aware. Table~\ref{tab:components} summarizes the components active in each configuration.

\begin{table}[tbp]
\centering
\caption{Components active in each configuration.}
\label{tab:components}
\begin{tabular}{lcc}
\toprule
\textbf{Config} & \textbf{Fidelity routing} & \textbf{Mirroring} \\
\midrule
SABRE   & \texttimes & \texttimes \\
FASST   & \checkmark & \texttimes \\
MIRAGE  & \texttimes & \checkmark \\
FINESSE & \checkmark & \checkmark \\
\bottomrule
\end{tabular}
\end{table}

These fidelity-aware approaches extend the architecture description---a weighted graph of coupling pairs---to incorporate gate fidelities $C_{ij}$ for each qubit pair:
\begin{equation}
L_{ij} = -\log \max(C_{ij},\, 10^{-10})
\end{equation}

As the circuit DAG is mapped onto the architecture, logical qubits are assigned to physical qubits, and each edge $(i,j)$ representing a placed gate is weighted by the corresponding $L_{ij}$. To capture the gate mapping impacts including downstream fidelity effects, the routing heuristic $H$ is modified to evaluate fidelity-weighted shortest paths over the circuit. Thus, by evaluating gates downstream using weights, i.e., $L_{ij}$, this $H'$ is implicitly evaluating both the infidelity of decomposed gates and those of necessary \sw{} operations on a path from $g$ to $g'$ due to executing the gates on the architecture topology. In particular, for each downstream gate $g' \in E$, paths from the current gate $g$ to $g'$ are evaluated using weights derived from the hardware cost of each intervening gate under the current mapping. These weights reflect the decomposition of gates into native operations and any required \sw{} operations imposed by connectivity constraints.

In SABRE, as fidelities are assumed to be uniform this simplifies to the proxy of hop counts $D$. In contrast, fidelity-aware approaches replace this with a blended distance $D'$ combining hop-count and fidelity-weighted shortest-path distances:
\begin{equation}
D'[i,j] = D_{\mathrm{hop}}[i,j] + \beta \cdot D_{\mathrm{fid}}[i,j]
\end{equation}
where $D_{\mathrm{fid}}[i,j]$ is the minimum accumulated log-infidelity path from $i$ to $j$ under Dijkstra with edge weights $k(\mathrm{SWAP},\mathcal{B})\cdot L_{ij}$, reflecting that each routing hop incurs a SWAP costing $k(\mathrm{SWAP},\mathcal{B})$ native basis gates for the basis gate set $\mathcal{B}$. The parameter $\beta \geq 0$ controls the relative weight of fidelity versus hop count; $\beta=0$ reduces the routing heuristic to hop-count distance, though FINESSE still applies fidelity-weighted mirror acceptance and lf-cost post-selection. We use $\beta = 1$ (equal weighting) throughout our benchmark suite. The decay factor from SABRE is supported as a configuration option but omitted in our benchmarks; the release valve introduced in LightSABRE provides sufficient cycle prevention, and the sum over the front layer is unnormalized by $\frac{1}{|F|}$ to match LightSABRE's relative scoring. The modified lookahead heuristic over the front layer $F$ and extended set $E$ is then:
\begin{equation}
\label{eq:H_dist}
H'(\pi) =\sum_{g \in F} D'[\pi(g)]
  + W \cdot \frac{1}{|E|}\sum_{g' \in E} D'[\pi(g')],
\end{equation}
where $\pi(g) = (p_0, p_1)$ denotes the physical qubits of gate $g$ under layout $\pi$, and $W = 0.5$ is a tunable parameter~\cite{li2019tackling}. 



\subsubsection{FINESSE mirror acceptance} FINESSE adapts MIRAGE's aggression~2 mirror acceptance to account for routing impact. When a two-qubit gate $U$ on physical edge $(p_0, p_1)$ becomes routable, the mirror $U' = \mathrm{SWAP} \cdot U$ is accepted if:
\begin{equation}
\label{eq:mirror_accept}
H'(\pi') +k(U_{ij}')L_{ij}\leq H'(\pi)+k(U_{ij})L_{ij},\end{equation}
where 
$\pi'$ is the layout after the implicit \sw{}, i.e., with the permuted outputs. 
Unlike MIRAGE, which accepts a mirror whenever $k(U') \leq k(U)$ regardless of routing impact, FINESSE's criterion is routing-aware: by including the heuristic terms $H'(\pi')$ and $H'(\pi)$, the acceptance condition accounts for whether the implicit qubit permutation improves or worsens the layout for future gates, avoiding mirrors that reduce decomposition cost at the expense of downstream fidelity.

\subsubsection{Fidelity-Aware Post-Selection}

After running $N$ independent routing trials, FINESSE selects the trial minimizing total log-fidelity cost, where the log-fidelity cost of each gate is weighted by the decomposition cost:
\begin{equation}
\label{eq:post_select}
\tau^* = \arg\min_{\tau} \sum_{g \in \mathcal{C}_\tau}
 L_{i_g, j_g}k(g),
\end{equation}
extending MIRAGE's depth-based post-selection with the same fidelity metric used throughout the pipeline.

\begin{figure*}[t]
  \centering
  \includegraphics[width=\textwidth]{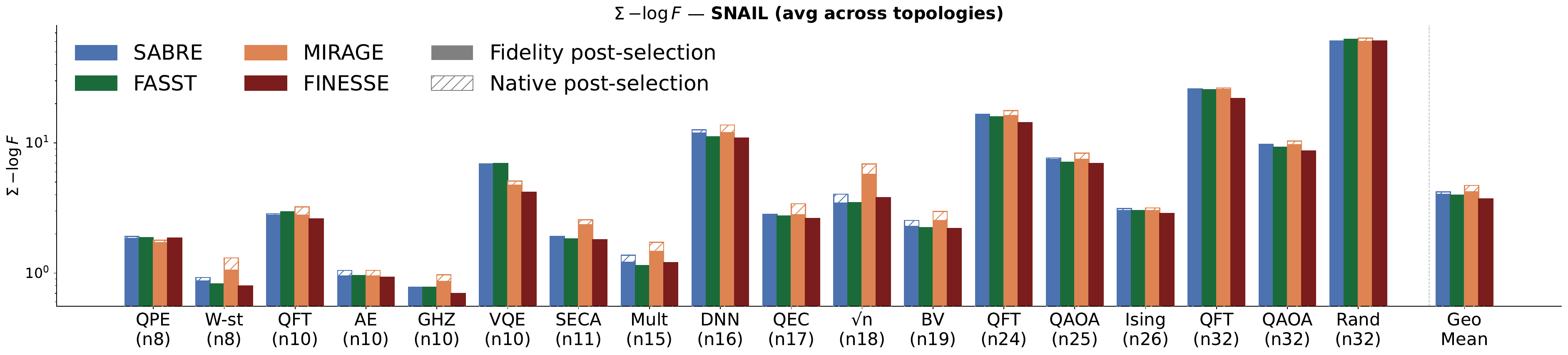}
  \caption{Total Log-infidelity of various circuits averaged over the results from all four topologies (lower is better)}
  \label{fig:lf_snail_avg}
\end{figure*}

\begin{figure*}[t]
  \centering
  \includegraphics[width=\textwidth]{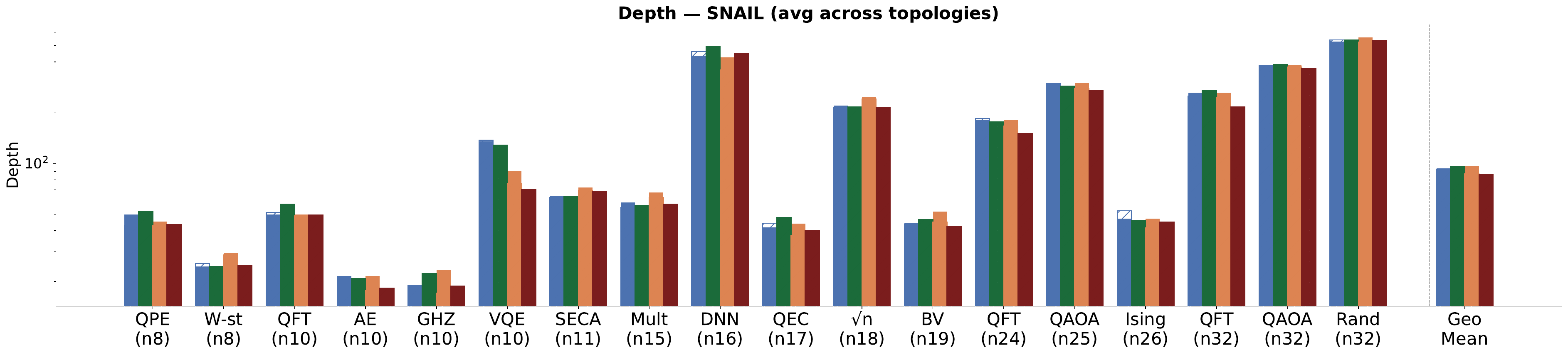}
  \caption{Total depth of various circuits averaged over the results from all four topologies (lower is better)}
  \label{fig:depth_snail_avg}
\end{figure*}

\section{Experimental Evaluation}
\label{sec:results}
To evaluate the impact of fidelity variation arising from frequency allocation in SNAIL-based architectures, the FASST and FINESSE algorithms were implemented as transpilation passes in Qiskit~\cite{qiskit}. Throughout this evaluation, our SABRE baseline employs LightSABRE's unnormalized heuristic and release valve; we refer to it as SABRE for brevity. The two-qubit decomposition costs $k(U)$ and $k(U')$ for mirror candidates are computed via Qiskit's Weyl chamber containment check, which determines the minimum number of native two-qubit gates required via KAK decomposition~\cite{tucci2005introduction} without performing explicit circuit decomposition. The resulting transpiled circuits are then verified using three methods. For circuits up to $8$ qubits, we compare the full $2^n \times 2^n$ unitary matrices directly. For circuits up to $15$ qubits, we evolve eight Haar-random input states through both the reference and routed circuits (with permutation correction) and confirm equivalence to tolerance $10^{-8}$. For Clifford circuits of any size, we additionally verify exact equivalence via Clifford tableau comparison. The full FINESSE pipeline can be found on Github.~\cite{finesse-repo}

\begin{table}[tbp]
\caption{Workloads from MQT Bench~\cite{quetschlich2023mqtbench} and QASMbench~\cite{li2020qasmbench}.}
    \centering
\begin{tabular}{l c|l c}
\toprule
\textbf{Circuit} & \textbf{Qubits} & \textbf{Circuit} & \textbf{Qubits}\\
\midrule
QPE                     & 8 &
QEC                     & 17\\
W-st                    & 8 &
$\sqrt{\text{n}}$       & 18\\
QFT                     & 10 &
BV                      & 19\\
AE                      & 10 &
QFT                     & 24\\
GHZ                     & 10 &
QAOA                    & 25\\
VQE                     & 10 &
Ising                   & 26\\
SECA                    & 11 &
QFT                     & 32\\
Mult                    & 15 &
QAOA                    & 32\\
DNN                     & 16 &
RAND                    & 32\\
\bottomrule
\end{tabular}
    \label{tab:workloads}
\end{table}

\begin{sloppypar}
We evaluate 18 quantum circuit workloads drawn from MQTBench~\cite{quetschlich2023mqtbench} and QASMbench~\cite{li2020qasmbench}, spanning 8 to 32 qubits as summarized in Table~\ref{tab:workloads}. The suite includes a mix of foundational algorithms (e.g., QPE, QFT, BV, AE), variational and optimization routines (QAOA, VQE, Ising), state preparation circuits (GHZ, W-state), arithmetic and learning kernels (multiplication, DNN), as well as random and error-correction circuits, capturing diverse structural patterns and application domains.  All circuits were run with 24 seeds and the best transpiled result was selected.  The workloads span both connectivity-constrained circuits with predominantly local interactions (e.g., QAOA, VQE, Ising, QEC) and end-to-end algorithms that induce global communication patterns (e.g., QFT, QPE, AE), enabling evaluation across routing-sensitive and long-range entanglement regimes.  This diversity captures the broad range of routing and fidelity trade-offs relevant to NISQ compilation. 
\end{sloppypar}

These workloads are evaluated on SNAIL-based architectures derived from Figures~\ref{fig:frequency_stack} and~\ref{fig:base-fidelities}, and extended to a larger fabric as shown in Figure~\ref{fig:corral}.

\begin{table}[tbp]
\centering
\caption{Two-qubit gate fidelities per edge (sorted highest to lowest) for each modular topology as reported in Figure~\ref{fig:base-fidelities}.}
\label{tab:topo_fidelities}
\resizebox{\columnwidth}{!}{%
\begin{tabular}{lcp{6cm}}
\hline
\textbf{Topology} & \textbf{Edges} & \textbf{Gate fidelities} \\
\hline
4Q, 4-edge module & 4 & 0.996,\ 0.995,\ 0.995,\ 0.994 \\
4Q, 5-edge module & 5 & 0.994,\ 0.993,\ 0.987,\ 0.986,\ 0.985 \\
4Q, 6-edge module & 6 & 0.994,\ 0.993,\ 0.991,\ 0.988,\ 0.977,\ 0.975 \\
5Q, 7-edge module & 7 & 0.989,\ 0.980,\ 0.973,\ 0.968,\ 0.968,\ 0.962,\ 0.960 \\
\hline
\end{tabular}%
}
\end{table}

\subsection{Multi-module Fabrics}
We evaluate four module topologies spanning 4--7 edges per module, with per-edge gate fidelities listed in Table~\ref{tab:topo_fidelities}. Three are four-qubit modules with 4, 5, and 6 edges; the fourth is a five-qubit module with 7 edges. As connectivity increases, frequency crowding degrades per-link fidelity, creating a tradeoff that the topology comparison makes explicit.

To demonstrate the effectiveness of FINESSE, Figures~\ref{fig:lf_snail_avg} and \ref{fig:depth_snail_avg} show the log-fidelity and depth, respectively, of MIRAGE, FASST, and FINESSE compared to SABRE, averaged over SNAIL topologies. We report results under two post-selection strategies. Under \emph{native post-selection}, each algorithm selects the seed that minimizes its own optimization objective: swap count for SABRE, circuit depth for MIRAGE, and log-infidelity cost for FASST and FINESSE. Under \emph{fidelity post-selection}, all algorithms select the seed with minimum log-infidelity cost. Native post-selection evaluates each transpiler as designed; fidelity post-selection provides a fair head-to-head comparison on the same metric. Mean results are summarized in Table~\ref{tab:results_summary}.

Generally FINESSE performs best, outperforming SABRE and MIRAGE in terms of fidelity. MIRAGE reduces depth through mirror gate absorption, but its depth-optimizing post-selection criterion tends to select seeds that sacrifice fidelity---explaining the increase in log-infidelity under fidelity post-selection, where that advantage is removed. The depth reduction for FINESSE is almost entirely attributable to mirror gates; the fidelity improvement comes from routing through higher-fidelity edges.

\begin{table}[tbp]
\centering
\caption{Mean \% change in log-infidelity cost and circuit depth vs.\ SABRE. Native post-selection: each algorithm minimizes its own objective. Fidelity post-selection: all algorithms select minimum log-infidelity seed.}
\label{tab:results_summary}
\resizebox{\columnwidth}{!}{%
\begin{tabular}{l rr rr}
\toprule
 & \multicolumn{2}{c}{\textbf{SNAIL (4 topologies)}} & \multicolumn{2}{c}{\textbf{IBM Brisbane}} \\
\cmidrule(lr){2-3}\cmidrule(lr){4-5}
\textbf{Config} & $\Delta$lf & $\Delta$depth & $\Delta$lf & $\Delta$depth \\
\midrule
\multicolumn{5}{l}{\emph{Native post-selection}} \\
FASST   & $-2.1\%$ & $+2.3\%$ & $-7.2\%$ & $+1.9\%$ \\
MIRAGE  & $+6.6\%$ & $-6.9\%$ & $-0.3\%$ & $-11.2\%$ \\
FINESSE & $-8.9\%$ & $-6.8\%$ & $-12.3\%$ & $-1.1\%$ \\
\midrule
\multicolumn{5}{l}{\emph{Fidelity post-selection}} \\
FASST   & $-0.5\%$ & $+2.6\%$ & $+3.0\%$ & $-0.6\%$ \\
MIRAGE  & $+1.3\%$ & $+1.2\%$ & $+0.4\%$ & $-0.1\%$ \\
FINESSE & $-7.4\%$ & $-6.5\%$ & $-2.6\%$ & $-3.5\%$ \\
\bottomrule
\end{tabular}%
}
\end{table}

\begin{figure*}[t]
  \centering
  \includegraphics[width=\textwidth]{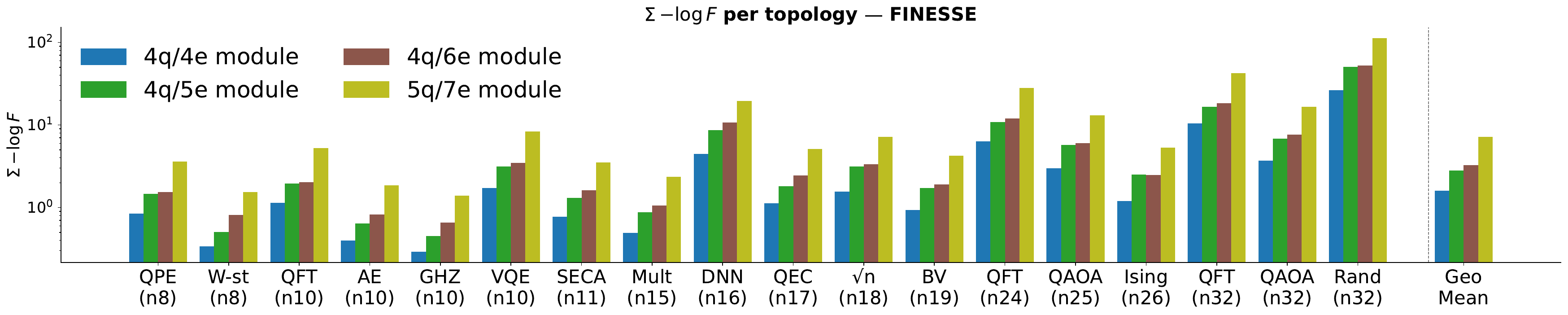}
  \caption{Log infidelity cost for each SNAIL-based topology with 4--7 edges per module using FINESSE transpiler (see Table~\ref{tab:topo_fidelities})}
  \label{fig:topo_lf_finesse_native}
\end{figure*}

Figure~\ref{fig:topo_lf_finesse_native} reports the average log-infidelity cost across circuits for each topology under FINESSE with fidelity post-selection. As module size and edge count increase, per-edge gate fidelities degrade (Table~\ref{tab:topo_fidelities}): the worst-performing link in the 4-edge module has fidelity 0.994, while the 7-edge module's worst link falls to 0.960. This degradation is a direct consequence of frequency crowding, as analyzed in Section~\ref{sec:allocation}: packing more qubits and couplers into a module forces qubit frequencies closer together, amplifying spectator-induced ZZ interactions and raising gate infidelity on the most constrained links.

Despite the higher connectivity of the 5- and 6-edge 4Q modules and the 7-edge 5Q module---which shortens routing paths and reduces SWAP overhead---these advantages are outweighed by the cost of executing on lower-fidelity links. The 4Q 4-edge module consistently achieves the lowest log-infidelity cost, confirming that in this regime link quality rather than connectivity density is the dominant factor in transpiled circuit fidelity. This presents a clear fidelity--connectivity tradeoff: beyond a threshold module size, each additional coupling edge introduces more fidelity overhead than it saves in routing cost, suggesting that modest connectivity with high per-link fidelity is preferable to dense connectivity with degraded links for NISQ-scale benchmarks.

\subsection{Generality for Fidelity-Aware Transpilation}
\begin{figure*}[tbp]
  \centering
  \includegraphics[width=\textwidth]{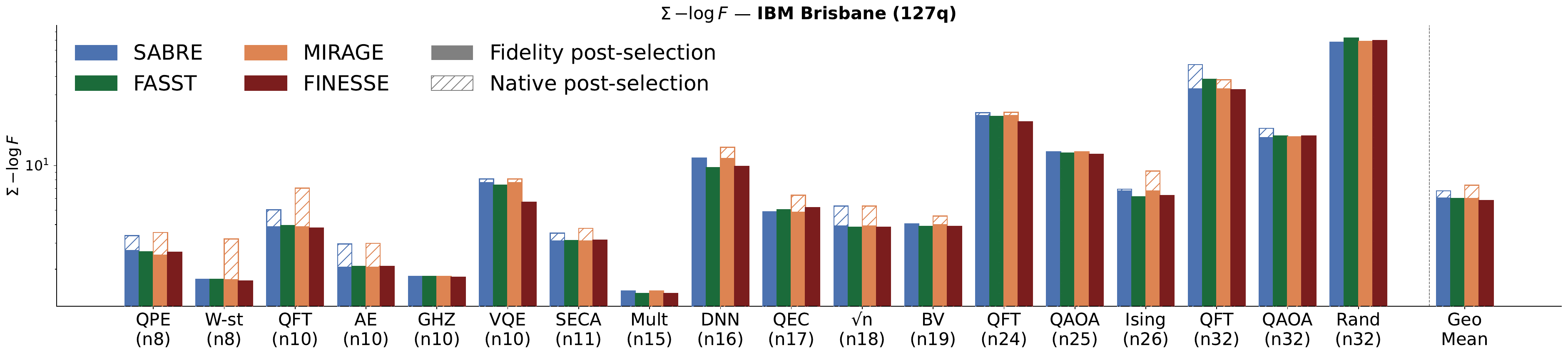}
  \caption{Log-infidelity cost for all transpilers on IBM's Brisbane device (127q, heavy-hex topology), under both post-selection methods.}
  \label{fig:ibm_brisbane}
\end{figure*}

FINESSE generalizes beyond frequency-allocation-driven noise models and can be applied to hardware platforms with heterogeneous gate fidelities arising from a variety of error sources. To demonstrate this, we present transpilation results on IBM's Brisbane device (127 qubits, heavy-hex topology), which implements two-qubit interactions via echoed cross-resonance (ECR) gates; results are reported in Figure~\ref{fig:ibm_brisbane}.

Per-link fidelity is derived directly from backend calibration data from IBM's FakeBrisbaneV2 provider, which contains a snapshot of real error rates for each connected qubit pair:
\begin{equation}
\label{eq:fidelity_conversion_ecr}
F_{ij} = 1 - e_{ij}^{\mathrm{ECR}},
\end{equation}
where $e_{ij}^{\mathrm{ECR}}$ denotes the calibrated \texttt{ECR} error rate on link $(i,j)$.

As shown in Table~\ref{tab:results_summary} and Figure~\ref{fig:ibm_brisbane}, FINESSE achieves a 12.3\% reduction in log-infidelity cost over SABRE under native post-selection, narrowing to 2.6\% under fidelity post-selection. Much of the native post-selection gap is attributable to SABRE's min-swap selection criterion: on Brisbane's 127-qubit heavy-hex topology, where ECR error rates vary substantially across links, seeds chosen to minimize swap count frequently route through low-fidelity edges. Fidelity post-selection allows SABRE to recover much of this loss, reducing FINESSE's remaining advantage. The larger gains under native post-selection on IBM compared to SNAIL reflect Brisbane's richer fidelity signal---the wider spread of link error rates gives the $D'$ heuristic more opportunity to redirect routing toward high-fidelity edges, consistent with FASST alone achieving 7.2\% on IBM versus 2.1\% on SNAIL.

These results confirm that the fidelity-weighted routing framework generalizes across native gate sets and topology structures, transferring directly to IBM's ECR-based heavy-hex hardware with no algorithm modification.

\section{Conclusion}
\label{sec:Conclusions}


The frequency allocation of qubits and conversion frequencies in SNAIL-based modules introduces a fundamental tradeoff between routing flexibility and gate fidelity. While modular architectures enable increased connectivity, scaling beyond four qubits and four coupling edges leads to significant frequency crowding, degrading two-qubit gate fidelity.

Fidelity-aware transpilation using FINESSE improves the utilization of available connectivity by preferentially routing through higher-fidelity interactions. However, these gains are bounded by underlying hardware constraints. In particular, the increase in infidelity from four- to five-qubit modules outweighs routing benefits, even under noise-aware compilation. This indicates that beyond a certain connectivity threshold, architectural degradation cannot be compensated for at the compiler level.

This work identifies a key limiting factor in modular system performance: the proximity of SNAIL, qubit, and subharmonic resonance frequencies. In particular, experimentally demonstrated limits on SNAIL operating frequencies constrain achievable frequency separation and exacerbate spectator interactions in larger qubit modules~\cite{2023APSMARG70003M}. Expanding the accessible SNAIL frequency range may enable improved fidelity and allow higher-connectivity modules to realize their theoretical routing advantages.


\bibliographystyle{ACM-Reference-Format}
\bibliography{Biblio/evan-refs,Biblio/mirage-refs,Biblio/dylan-refs}


\begin{thebibliography}{48}


\ifx \showCODEN    \undefined \def \showCODEN     #1{\unskip}     \fi
\ifx \showISBNx    \undefined \def \showISBNx     #1{\unskip}     \fi
\ifx \showISBNxiii \undefined \def \showISBNxiii  #1{\unskip}     \fi
\ifx \showISSN     \undefined \def \showISSN      #1{\unskip}     \fi
\ifx \showLCCN     \undefined \def \showLCCN      #1{\unskip}     \fi
\ifx \shownote     \undefined \def \shownote      #1{#1}          \fi
\ifx \showarticletitle \undefined \def \showarticletitle #1{#1}   \fi
\ifx \showURL      \undefined \def \showURL       {\relax}        \fi
\providecommand\bibfield[2]{#2}
\providecommand\bibinfo[2]{#2}
\providecommand\natexlab[1]{#1}
\providecommand\showeprint[2][]{arXiv:#2}

\bibitem[Jones et~al\mbox{.}(2012)]%
        {jones_layered_2012}
\bibfield{author}{\bibinfo{person}{N~Cody Jones}, \bibinfo{person}{Rodney
  Van~Meter}, \bibinfo{person}{Austin~G Fowler}, \bibinfo{person}{Peter~L
  McMahon}, \bibinfo{person}{Jungsang Kim}, \bibinfo{person}{Thaddeus~D Ladd},
  {and} \bibinfo{person}{Yoshihisa Yamamoto}.} \bibinfo{year}{2012}\natexlab{}.
\newblock \showarticletitle{Layered architecture for quantum computing}.
\newblock \bibinfo{journal}{\emph{Phys.Rev.X}} (\bibinfo{year}{2012}).
\newblock
\href{https://doi.org/10.1103/PhysRevX.2.031007}{doi:\nolinkurl{10.1103/PhysRevX.2.031007}}


\bibitem[McKinney et~al\mbox{.}(2023)]%
        {mckinney_co-designed_2023}
\bibfield{author}{\bibinfo{person}{Evan McKinney}, \bibinfo{person}{Mingkang
  Xia}, \bibinfo{person}{Chao Zhou}, \bibinfo{person}{Pinlei Lu},
  \bibinfo{person}{Michael Hatridge}, {and} \bibinfo{person}{Alex~K Jones}.}
  \bibinfo{year}{2023}\natexlab{}.
\newblock \showarticletitle{Co-designed architectures for modular
  superconducting quantum computers}. In \bibinfo{booktitle}{\emph{2023 IEEE
  International Symposium on High-Performance Computer Architecture (HPCA)}}.
  IEEE, \bibinfo{pages}{759--772}.
\newblock


\bibitem[Beverland et~al\mbox{.}(2022)]%
        {beverland_assessing_2022}
\bibfield{author}{\bibinfo{person}{Michael~E Beverland},
  \bibinfo{person}{Prakash Murali}, \bibinfo{person}{Matthias Troyer},
  \bibinfo{person}{Krysta~M Svore}, \bibinfo{person}{Torsten Hoefler},
  \bibinfo{person}{Vadym Kliuchnikov}, \bibinfo{person}{Guang~Hao Low},
  \bibinfo{person}{Mathias Soeken}, \bibinfo{person}{Aarthi Sundaram}, {and}
  \bibinfo{person}{Alexander Vaschillo}.} \bibinfo{year}{2022}\natexlab{}.
\newblock \showarticletitle{Assessing requirements to scale to practical
  quantum advantage}.
\newblock \bibinfo{journal}{\emph{arXiv preprint arXiv:2211.07629}}
  (\bibinfo{year}{2022}).
\newblock


\bibitem[Tomesh and Martonosi(2021)]%
        {tomesh_quantum_2021}
\bibfield{author}{\bibinfo{person}{Teague Tomesh} {and}
  \bibinfo{person}{Margaret Martonosi}.} \bibinfo{year}{2021}\natexlab{}.
\newblock \showarticletitle{Quantum codesign}.
\newblock \bibinfo{journal}{\emph{IEEE Micro}} \bibinfo{volume}{41},
  \bibinfo{number}{5} (\bibinfo{year}{2021}), \bibinfo{pages}{33--40}.
\newblock


\bibitem[Murali et~al\mbox{.}(2020)]%
        {murali2020software}
\bibfield{author}{\bibinfo{person}{Prakash Murali}, \bibinfo{person}{David~C
  McKay}, \bibinfo{person}{Margaret Martonosi}, {and} \bibinfo{person}{Ali
  Javadi-Abhari}.} \bibinfo{year}{2020}\natexlab{}.
\newblock \showarticletitle{Software mitigation of crosstalk on noisy
  intermediate-scale quantum computers}. In
  \bibinfo{booktitle}{\emph{Proceedings of the Twenty-Fifth International
  Conference on Architectural Support for Programming Languages and Operating
  Systems}}. \bibinfo{pages}{1001--1016}.
\newblock


\bibitem[Murali et~al\mbox{.}(2019)]%
        {murali_full-stack_2019}
\bibfield{author}{\bibinfo{person}{Prakash Murali},
  \bibinfo{person}{Norbert~Matthias Linke}, \bibinfo{person}{Margaret
  Martonosi}, \bibinfo{person}{Ali~Javadi Abhari}, \bibinfo{person}{Nhung~Hong
  Nguyen}, {and} \bibinfo{person}{Cinthia~Huerta Alderete}.}
  \bibinfo{year}{2019}\natexlab{}.
\newblock \showarticletitle{Full-stack, real-system quantum computer studies:
  Architectural comparisons and design insights}. In
  \bibinfo{booktitle}{\emph{Proceedings of the 46th International Symposium on
  Computer Architecture}}. \bibinfo{pages}{527--540}.
\newblock


\bibitem[Zhou et~al\mbox{.}(2023)]%
        {zhou_realizing_2023}
\bibfield{author}{\bibinfo{person}{Chao Zhou}, \bibinfo{person}{Pinlei Lu},
  \bibinfo{person}{Matthieu Praquin}, \bibinfo{person}{Tzu-Chiao Chien},
  \bibinfo{person}{Ryan Kaufman}, \bibinfo{person}{Xi Cao},
  \bibinfo{person}{Mingkang Xia}, \bibinfo{person}{Roger~SK Mong},
  \bibinfo{person}{Wolfgang Pfaff}, \bibinfo{person}{David Pekker},
  {et~al\mbox{.}}} \bibinfo{year}{2023}\natexlab{}.
\newblock \showarticletitle{Realizing all-to-all couplings among detachable
  quantum modules using a microwave quantum state router}.
\newblock \bibinfo{journal}{\emph{npj quantum information}}
  \bibinfo{volume}{9}, \bibinfo{number}{1} (\bibinfo{year}{2023}),
  \bibinfo{pages}{54}.
\newblock


\bibitem[Brink et~al\mbox{.}(2018)]%
        {brink_device_2018}
\bibfield{author}{\bibinfo{person}{Markus Brink}, \bibinfo{person}{Jerry~M.
  Chow}, \bibinfo{person}{Jared Hertzberg}, \bibinfo{person}{Easwar Magesan},
  {and} \bibinfo{person}{Sami Rosenblatt}.} \bibinfo{year}{2018}\natexlab{}.
\newblock \showarticletitle{Device challenges for near term superconducting
  quantum processors: frequency collisions}. In
  \bibinfo{booktitle}{\emph{IEDM}}.
\newblock
\showISBNx{978-1-72811-987-8}
\href{https://doi.org/10.1109/IEDM.2018.8614500}{doi:\nolinkurl{10.1109/IEDM.2018.8614500}}


\bibitem[Tripathi et~al\mbox{.}(2019)]%
        {tripathi_operation_2019}
\bibfield{author}{\bibinfo{person}{Vinay Tripathi}, \bibinfo{person}{Mostafa
  Khezri}, {and} \bibinfo{person}{Alexander~N Korotkov}.}
  \bibinfo{year}{2019}\natexlab{}.
\newblock \showarticletitle{Operation and intrinsic error budget of a two-qubit
  cross-resonance gate}.
\newblock \bibinfo{journal}{\emph{Physical Review A}} \bibinfo{volume}{100},
  \bibinfo{number}{1} (\bibinfo{year}{2019}), \bibinfo{pages}{012301}.
\newblock


\bibitem[Ni et~al\mbox{.}(2023)]%
        {ni_superconducting_2023}
\bibfield{author}{\bibinfo{person}{Xiaotong Ni}, \bibinfo{person}{Ziang Wang},
  \bibinfo{person}{Rui Chao}, {and} \bibinfo{person}{Jianxin Chen}.}
  \bibinfo{year}{2023}\natexlab{}.
\newblock \showarticletitle{Superconducting processor design optimization for
  quantum error correction performance}.
\newblock \bibinfo{journal}{\emph{arXiv preprint arXiv:2312.04186}}
  (\bibinfo{year}{2023}).
\newblock


\bibitem[Sete et~al\mbox{.}(2024)]%
        {sete_error_2024}
\bibfield{author}{\bibinfo{person}{Eyob~A Sete}, \bibinfo{person}{Vinay
  Tripathi}, \bibinfo{person}{Joseph~A Valery}, \bibinfo{person}{Daniel Lidar},
  {and} \bibinfo{person}{Josh~Y Mutus}.} \bibinfo{year}{2024}\natexlab{}.
\newblock \showarticletitle{Error budget of a parametric resonance entangling
  gate with a tunable coupler}.
\newblock \bibinfo{journal}{\emph{Physical Review Applied}}
  \bibinfo{volume}{22}, \bibinfo{number}{1} (\bibinfo{year}{2024}),
  \bibinfo{pages}{014059}.
\newblock


\bibitem[Li et~al\mbox{.}(2020)]%
        {li_towards_2019}
\bibfield{author}{\bibinfo{person}{Gushu Li}, \bibinfo{person}{Yufei Ding},
  {and} \bibinfo{person}{Yuan Xie}.} \bibinfo{year}{2020}\natexlab{}.
\newblock \showarticletitle{Towards efficient superconducting quantum processor
  architecture design}. In \bibinfo{booktitle}{\emph{Proceedings of the
  Twenty-Fifth International Conference on Architectural Support for
  Programming Languages and Operating Systems}}. \bibinfo{pages}{1031--1045}.
\newblock


\bibitem[Ding et~al\mbox{.}(2020)]%
        {ding_systematic_2020}
\bibfield{author}{\bibinfo{person}{Yongshan Ding}, \bibinfo{person}{Pranav
  Gokhale}, \bibinfo{person}{Sophia~Fuhui Lin}, \bibinfo{person}{Richard
  Rines}, \bibinfo{person}{Thomas Propson}, {and} \bibinfo{person}{Frederic~T
  Chong}.} \bibinfo{year}{2020}\natexlab{}.
\newblock \showarticletitle{Systematic crosstalk mitigation for superconducting
  qubits via frequency-aware compilation}. In \bibinfo{booktitle}{\emph{2020
  53rd Annual IEEE/ACM International Symposium on Microarchitecture (MICRO)}}.
  IEEE, \bibinfo{pages}{201--214}.
\newblock


\bibitem[Smith et~al\mbox{.}(2022)]%
        {smith_scaling_2022}
\bibfield{author}{\bibinfo{person}{Kaitlin~N Smith},
  \bibinfo{person}{Gokul~Subramanian Ravi}, \bibinfo{person}{Jonathan~M Baker},
  {and} \bibinfo{person}{Frederic~T Chong}.} \bibinfo{year}{2022}\natexlab{}.
\newblock \showarticletitle{Scaling superconducting quantum computers with
  chiplet architectures}. In \bibinfo{booktitle}{\emph{2022 55th IEEE/ACM
  International Symposium on Microarchitecture (MICRO)}}. IEEE,
  \bibinfo{pages}{1092--1109}.
\newblock


\bibitem[Morvan et~al\mbox{.}(2022)]%
        {morvan_optimizing_2022}
\bibfield{author}{\bibinfo{person}{Alexis Morvan}, \bibinfo{person}{Larry
  Chen}, \bibinfo{person}{Jeffrey~M. Larson}, \bibinfo{person}{David~I.
  Santiago}, {and} \bibinfo{person}{Irfan Siddiqi}.}
  \bibinfo{year}{2022}\natexlab{}.
\newblock \showarticletitle{Optimizing frequency allocation for fixed-frequency
  superconducting quantum processors}.
\newblock \bibinfo{journal}{\emph{Phys.Rev.Res.}} (\bibinfo{date}{April}
  \bibinfo{year}{2022}).
\newblock
\href{https://doi.org/10.1103/PhysRevResearch.4.023079}{doi:\nolinkurl{10.1103/PhysRevResearch.4.023079}}


\bibitem[Osman et~al\mbox{.}(2023)]%
        {osman_mitigation_2023}
\bibfield{author}{\bibinfo{person}{Amr Osman}, \bibinfo{person}{Jorge
  Fernández-Pendás}, \bibinfo{person}{Christopher Warren},
  \bibinfo{person}{Sandoko Kosen}, \bibinfo{person}{Marco Scigliuzzo},
  \bibinfo{person}{Anton Frisk~Kockum}, \bibinfo{person}{Giovanna Tancredi},
  \bibinfo{person}{Anita Fadavi~Roudsari}, {and} \bibinfo{person}{Jonas
  Bylander}.} \bibinfo{year}{2023}\natexlab{}.
\newblock \showarticletitle{Mitigation of frequency collisions in
  superconducting quantum processors}.
\newblock \bibinfo{journal}{\emph{Phys.Rev.Res.}} (\bibinfo{date}{Oct.}
  \bibinfo{year}{2023}).
\newblock
\href{https://doi.org/10.1103/PhysRevResearch.5.043001}{doi:\nolinkurl{10.1103/PhysRevResearch.5.043001}}


\bibitem[Zhang et~al\mbox{.}(2024)]%
        {zhang_qplacer_2024}
\bibfield{author}{\bibinfo{person}{Junyao Zhang}, \bibinfo{person}{Hanrui
  Wang}, \bibinfo{person}{Qi Ding}, \bibinfo{person}{Jiaqi Gu},
  \bibinfo{person}{Reouven Assouly}, \bibinfo{person}{William~D Oliver},
  \bibinfo{person}{Song Han}, \bibinfo{person}{Kenneth~R Brown},
  \bibinfo{person}{Hai Li}, \bibinfo{person}{Yiran Chen}, {et~al\mbox{.}}}
  \bibinfo{year}{2024}\natexlab{}.
\newblock \showarticletitle{Qplacer: Frequency-aware component placement for
  superconducting quantum computers}.
\newblock \bibinfo{journal}{\emph{arXiv preprint arXiv:2401.17450}}
  (\bibinfo{year}{2024}).
\newblock


\bibitem[Zhang et~al\mbox{.}(2025)]%
        {zhangEfficientFrequencyAllocation2024}
\bibfield{author}{\bibinfo{person}{Zewen Zhang}, \bibinfo{person}{Pranav
  Gokhale}, {and} \bibinfo{person}{Jeffrey~M Larson}.}
  \bibinfo{year}{2025}\natexlab{}.
\newblock \showarticletitle{Efficient frequency allocation for superconducting
  quantum processors using improved optimization techniques}.
\newblock \bibinfo{journal}{\emph{Physical Review A}} \bibinfo{volume}{111},
  \bibinfo{number}{1} (\bibinfo{year}{2025}), \bibinfo{pages}{012619}.
\newblock


\bibitem[Li et~al\mbox{.}(2019)]%
        {li2019tackling}
\bibfield{author}{\bibinfo{person}{Gushu Li}, \bibinfo{person}{Yufei Ding},
  {and} \bibinfo{person}{Yuan Xie}.} \bibinfo{year}{2019}\natexlab{}.
\newblock \showarticletitle{Tackling the qubit mapping problem for NISQ-era
  quantum devices}. In \bibinfo{booktitle}{\emph{Proceedings of the
  Twenty-Fourth International Conference on Architectural Support for
  Programming Languages and Operating Systems}}. \bibinfo{pages}{1001--1014}.
\newblock


\bibitem[McKinney et~al\mbox{.}(2024)]%
        {mckinney2024mirage}
\bibfield{author}{\bibinfo{person}{Evan McKinney}, \bibinfo{person}{Michael
  Hatridge}, {and} \bibinfo{person}{Alex~K Jones}.}
  \bibinfo{year}{2024}\natexlab{}.
\newblock \showarticletitle{MIRAGE: Quantum circuit decomposition and routing
  collaborative design using mirror gates}. In \bibinfo{booktitle}{\emph{2024
  IEEE International Symposium on High-Performance Computer Architecture
  (HPCA)}}. IEEE, \bibinfo{pages}{704--718}.
\newblock


\bibitem[Frattini et~al\mbox{.}(2017)]%
        {frattini17}
\bibfield{author}{\bibinfo{person}{N.E. Frattini}, \bibinfo{person}{U. Vool},
  \bibinfo{person}{A. Narla}, \bibinfo{person}{K.M. Swila}, {and}
  \bibinfo{person}{M.H. Devoret}.} \bibinfo{year}{2017}\natexlab{}.
\newblock \showarticletitle{3-wave mixing Josephson dipole element}.
\newblock \bibinfo{journal}{\emph{Applied physics letters}}
  \bibinfo{volume}{110}, \bibinfo{number}{222603} (\bibinfo{year}{2017}).
\newblock


\bibitem[Koch et~al\mbox{.}(2007)]%
        {Koch07}
\bibfield{author}{\bibinfo{person}{Jens Koch}, \bibinfo{person}{M.~Yul Terri},
  \bibinfo{person}{Jay Gambetta}, \bibinfo{person}{A.A. Houck},
  \bibinfo{person}{D.I. Schuster}, \bibinfo{person}{J. Majer},
  \bibinfo{person}{Alexandre Blais}, \bibinfo{person}{M.H. Devoret},
  \bibinfo{person}{S.M. Girvin}, {et~al\mbox{.}}}
  \bibinfo{year}{2007}\natexlab{}.
\newblock \showarticletitle{Charge-insensitive qubit design derived from the
  Cooper pair box}.
\newblock \bibinfo{journal}{\emph{Physical Review Letters A}}
  (\bibinfo{year}{2007}).
\newblock


\bibitem[Nielsen(2002)]%
        {nielsen_simple_2002}
\bibfield{author}{\bibinfo{person}{Michael~A. Nielsen}.}
  \bibinfo{year}{2002}\natexlab{}.
\newblock \showarticletitle{A simple formula for the average gate fidelity of a
  quantum dynamical operation}.
\newblock \bibinfo{journal}{\emph{Phys. Lett. A}} (\bibinfo{date}{Oct.}
  \bibinfo{year}{2002}), \bibinfo{pages}{249--252}.
\newblock
\href{https://doi.org/10.1016/S0375-9601(02)01272-0}{doi:\nolinkurl{10.1016/S0375-9601(02)01272-0}}


\bibitem[Hopf et~al\mbox{.}(2025)]%
        {hopf2025improving}
\bibfield{author}{\bibinfo{person}{Patrick Hopf}, \bibinfo{person}{Nils
  Quetschlich}, \bibinfo{person}{Laura Schulz}, {and} \bibinfo{person}{Robert
  Wille}.} \bibinfo{year}{2025}\natexlab{}.
\newblock \showarticletitle{Improving Figures of Merit for Quantum Circuit
  Compilation}.
\newblock \bibinfo{journal}{\emph{arXiv preprint arXiv:2501.13155}}
  (\bibinfo{year}{2025}).
\newblock


\bibitem[Gokhale et~al\mbox{.}(2024)]%
        {gokhale2024faster}
\bibfield{author}{\bibinfo{person}{Pranav Gokhale}, \bibinfo{person}{Teague
  Tomesh}, \bibinfo{person}{Martin Suchara}, {and} \bibinfo{person}{Fred
  Chong}.} \bibinfo{year}{2024}\natexlab{}.
\newblock \showarticletitle{Faster and more reliable quantum swaps via native
  gates}. In \bibinfo{booktitle}{\emph{Proceedings of the 2024 International
  Conference on Parallel Architectures and Compilation Techniques}}.
  \bibinfo{pages}{351--362}.
\newblock


\bibitem[Schmid et~al\mbox{.}(2024)]%
        {schmid2024computational}
\bibfield{author}{\bibinfo{person}{Ludwig Schmid}, \bibinfo{person}{David~F
  Locher}, \bibinfo{person}{Manuel Rispler}, \bibinfo{person}{Sebastian Blatt},
  \bibinfo{person}{Johannes Zeiher}, \bibinfo{person}{Markus M{\"u}ller}, {and}
  \bibinfo{person}{Robert Wille}.} \bibinfo{year}{2024}\natexlab{}.
\newblock \showarticletitle{Computational capabilities and compiler development
  for neutral atom quantum processors—connecting tool developers and hardware
  experts}.
\newblock \bibinfo{journal}{\emph{Quantum Science and Technology}}
  \bibinfo{volume}{9}, \bibinfo{number}{3} (\bibinfo{year}{2024}),
  \bibinfo{pages}{033001}.
\newblock


\bibitem[Chen et~al\mbox{.}(2024)]%
        {chen_one_2023}
\bibfield{author}{\bibinfo{person}{Jianxin Chen}, \bibinfo{person}{Dawei Ding},
  \bibinfo{person}{Weiyuan Gong}, \bibinfo{person}{Cupjin Huang}, {and}
  \bibinfo{person}{Qi Ye}.} \bibinfo{year}{2024}\natexlab{}.
\newblock \showarticletitle{One Gate Scheme to Rule Them All: Introducing a
  Complex Yet Reduced Instruction Set for Quantum Computing}. In
  \bibinfo{booktitle}{\emph{ASPLOS}}.
\newblock
\showISBNx{9798400703850}
\href{https://doi.org/10.1145/3620665.3640386}{doi:\nolinkurl{10.1145/3620665.3640386}}


\bibitem[Zhou(2023)]%
        {zhou_superconducting_2023}
\bibfield{author}{\bibinfo{person}{Chao Zhou}.}
  \bibinfo{year}{2023}\natexlab{}.
\newblock \emph{\bibinfo{title}{Superconducting {Quantum} {Routers}, {Modules},
  {Gates}, and {Measurements} {Based} on {Charge}-pumped {Parametric}
  {Interactions}}}.
\newblock \bibinfo{thesistype}{Ph.\,D. Dissertation}.
  \bibinfo{school}{University of Pittsburgh}.
\newblock


\bibitem[Xia et~al\mbox{.}(2023)]%
        {xia2023fast}
\bibfield{author}{\bibinfo{person}{Mingkang Xia}, \bibinfo{person}{Chao Zhou},
  \bibinfo{person}{Chenxu Liu}, \bibinfo{person}{Param Patel},
  \bibinfo{person}{Xi Cao}, \bibinfo{person}{Pinlei Lu}, \bibinfo{person}{Boris
  Mesits}, \bibinfo{person}{Maria Mucci}, \bibinfo{person}{David Gorski},
  \bibinfo{person}{David Pekker}, {et~al\mbox{.}}}
  \bibinfo{year}{2023}\natexlab{}.
\newblock \showarticletitle{Fast superconducting qubit control with
  sub-harmonic drives}.
\newblock \bibinfo{journal}{\emph{arXiv preprint arXiv:2306.10162}}
  (\bibinfo{year}{2023}).
\newblock


\bibitem[Barajas and Campbell(2025)]%
        {barajas2025quantum}
\bibfield{author}{\bibinfo{person}{Kristian~D Barajas} {and}
  \bibinfo{person}{Wesley~C Campbell}.} \bibinfo{year}{2025}\natexlab{}.
\newblock \showarticletitle{Quantum Averaging for High-Fidelity Quantum Logic
  Gates}.
\newblock \bibinfo{journal}{\emph{arXiv preprint arXiv:2503.08886}}
  (\bibinfo{year}{2025}).
\newblock


\bibitem[McKinney et~al\mbox{.}(2023)]%
        {mckinney2023parallel}
\bibfield{author}{\bibinfo{person}{Evan McKinney}, \bibinfo{person}{Chao Zhou},
  \bibinfo{person}{Mingkang Xia}, \bibinfo{person}{Michael Hatridge}, {and}
  \bibinfo{person}{Alex~K Jones}.} \bibinfo{year}{2023}\natexlab{}.
\newblock \showarticletitle{Parallel driving for fast quantum computing under
  speed limits}. In \bibinfo{booktitle}{\emph{Proceedings of the 50th Annual
  International Symposium on Computer Architecture}}. \bibinfo{pages}{1--13}.
\newblock


\bibitem[Dumas et~al\mbox{.}(2024)]%
        {dumas2024unified}
\bibfield{author}{\bibinfo{person}{Marie~Fr\'ed\'erique Dumas},
  \bibinfo{person}{Benjamin Groleau-Par\'e}, \bibinfo{person}{Alexander
  McDonald}, \bibinfo{person}{Manuel~H. Mu\~noz Arias},
  \bibinfo{person}{Crist\'obal Lled\'o}, \bibinfo{person}{Benjamin D'Anjou},
  {and} \bibinfo{person}{Alexandre Blais}.} \bibinfo{year}{2024}\natexlab{}.
\newblock \showarticletitle{{Measurement-Induced Transmon Ionization}}.
\newblock \bibinfo{journal}{\emph{Phys. Rev. X}} \bibinfo{volume}{14},
  \bibinfo{number}{4} (\bibinfo{year}{2024}), \bibinfo{pages}{041023}.
\newblock
\showeprint[arxiv]{2402.06615}~[quant-ph]
\href{https://doi.org/10.1103/PhysRevX.14.041023}{doi:\nolinkurl{10.1103/PhysRevX.14.041023}}


\bibitem[Frattini(2021)]%
        {frattini2021three}
\bibfield{author}{\bibinfo{person}{Nicholas~E Frattini}.}
  \bibinfo{year}{2021}\natexlab{}.
\newblock \emph{\bibinfo{title}{Three-wave mixing in superconducting circuits:
  stabilizing cats with SNAILs}}.
\newblock \bibinfo{thesistype}{Ph.\,D. Dissertation}. \bibinfo{school}{Yale
  University}.
\newblock


\bibitem[Larkin(2018)]%
        {Raman2018}
\bibfield{author}{\bibinfo{person}{Peter~J. Larkin}.}
  \bibinfo{year}{2018}\natexlab{}.
\newblock \bibinfo{booktitle}{\emph{Infrared and Raman Spectroscopy (Second
  Edition)}}.
\newblock \bibinfo{publisher}{Elsevier}.
\newblock


\bibitem[Park and Lee(1996)]%
        {park_application_1996}
\bibfield{author}{\bibinfo{person}{Taehoon Park} {and} \bibinfo{person}{Chae~Y.
  Lee}.} \bibinfo{year}{1996}\natexlab{}.
\newblock \showarticletitle{Application of the Graph Coloring Algorithm to the
  Frequency Assignment Problem}.
\newblock \bibinfo{journal}{\emph{Journal of the Operations Research Society of
  Japan}} (\bibinfo{year}{1996}).
\newblock
\showISSN{0453-4514, 2188-8299}
\href{https://doi.org/10.15807/jorsj.39.258}{doi:\nolinkurl{10.15807/jorsj.39.258}}


\bibitem[Waters(2005)]%
        {waters2005graph}
\bibfield{author}{\bibinfo{person}{Robert~James Waters}.}
  \bibinfo{year}{2005}\natexlab{}.
\newblock \emph{\bibinfo{title}{Graph colouring and frequency assignment}}.
\newblock \bibinfo{thesistype}{Ph.\,D. Dissertation}.
  \bibinfo{school}{University of Nottingham}.
\newblock


\bibitem[Orden et~al\mbox{.}(2018)]%
        {orden_spectrum_2018}
\bibfield{author}{\bibinfo{person}{David Orden}, \bibinfo{person}{Jose~Manuel
  Gimenez-Guzman}, \bibinfo{person}{Ivan Marsa-Maestre}, {and}
  \bibinfo{person}{Enrique De~la Hoz}.} \bibinfo{year}{2018}\natexlab{}.
\newblock \showarticletitle{Spectrum {Graph} {Coloring} and {Applications} to
  {Wi}-{Fi} {Channel} {Assignment}}.
\newblock \bibinfo{journal}{\emph{Symmetry}} (\bibinfo{date}{March}
  \bibinfo{year}{2018}).
\newblock
\href{https://doi.org/10.3390/sym10030065}{doi:\nolinkurl{10.3390/sym10030065}}


\bibitem[Memarsadeghi(2016)]%
        {memarsadeghi2016nasa}
\bibfield{author}{\bibinfo{person}{Nargess Memarsadeghi}.}
  \bibinfo{year}{2016}\natexlab{}.
\newblock \showarticletitle{NASA computational case study: Golomb rulers and
  their applications}.
\newblock \bibinfo{journal}{\emph{Computing in Science \& Engineering}}
  \bibinfo{volume}{18}, \bibinfo{number}{06} (\bibinfo{year}{2016}),
  \bibinfo{pages}{58--62}.
\newblock


\bibitem[Meyer and Papakonstantinou(2009)]%
        {meyer2009complexity}
\bibfield{author}{\bibinfo{person}{Christophe Meyer} {and}
  \bibinfo{person}{Periklis~A Papakonstantinou}.}
  \bibinfo{year}{2009}\natexlab{}.
\newblock \showarticletitle{On the complexity of constructing Golomb rulers}.
\newblock \bibinfo{journal}{\emph{Discrete applied mathematics}}
  \bibinfo{volume}{157}, \bibinfo{number}{4} (\bibinfo{year}{2009}),
  \bibinfo{pages}{738--748}.
\newblock


\bibitem[Ai and Liu(2024)]%
        {ai2024graph}
\bibfield{author}{\bibinfo{person}{Hao Ai} {and} \bibinfo{person}{Yu-xi Liu}.}
  \bibinfo{year}{2024}\natexlab{}.
\newblock \showarticletitle{Graph Neural Networks-based Parameter Design
  towards Large-Scale Superconducting Quantum Circuits for Crosstalk
  Mitigation}.
\newblock \bibinfo{journal}{\emph{arXiv preprint arXiv:2411.16354}}
  (\bibinfo{year}{2024}).
\newblock


\bibitem[Mammola et~al\mbox{.}(2025)]%
        {mammola2025optimal}
\bibfield{author}{\bibinfo{person}{Andrea Mammola}, \bibinfo{person}{Quentin
  Schaeverbeke}, {and} \bibinfo{person}{Matthieu~M Desjardins}.}
  \bibinfo{year}{2025}\natexlab{}.
\newblock \showarticletitle{Optimal Connectivity from Idle Qubit residual
  coupling Cross-Talks in a Cavity Mediated Entangling Gate}.
\newblock \bibinfo{journal}{\emph{arXiv preprint arXiv:2503.07455}}
  (\bibinfo{year}{2025}).
\newblock


\bibitem[Zou et~al\mbox{.}(2024)]%
        {Zou2024}
\bibfield{author}{\bibinfo{person}{Henry Zou}, \bibinfo{person}{Matthew
  Treinish}, \bibinfo{person}{Kevin Hartman}, \bibinfo{person}{Alexander
  Ivrii}, {and} \bibinfo{person}{Jake Lishman}.}
  \bibinfo{year}{2024}\natexlab{}.
\newblock \showarticletitle{{LightSABRE}: A Lightweight and Enhanced {SABRE}
  Algorithm}.
\newblock \bibinfo{journal}{\emph{arXiv preprint arXiv:2409.08368}}
  (\bibinfo{year}{2024}).
\newblock


\bibitem[{Qiskit contributors}(2023)]%
        {qiskit}
\bibfield{author}{\bibinfo{person}{{Qiskit contributors}}.}
  \bibinfo{year}{2023}\natexlab{}.
\newblock \bibinfo{title}{Qiskit: An Open-source Framework for Quantum
  Computing}.
\newblock
\href{https://doi.org/10.5281/zenodo.2573505}{doi:\nolinkurl{10.5281/zenodo.2573505}}


\bibitem[Tucci(2005)]%
        {tucci2005introduction}
\bibfield{author}{\bibinfo{person}{Robert~R Tucci}.}
  \bibinfo{year}{2005}\natexlab{}.
\newblock \showarticletitle{An introduction to Cartan's KAK decomposition for
  QC programmers}.
\newblock \bibinfo{journal}{\emph{arXiv preprint quant-ph/0507171}}
  (\bibinfo{year}{2005}).
\newblock


\bibitem[VanAllen(2026)]%
        {finesse-repo}
\bibfield{author}{\bibinfo{person}{Dylan VanAllen}.}
  \bibinfo{year}{2026}\natexlab{}.
\newblock \bibinfo{title}{FINESSE: Fidelity-Integrated Equivalence-Aware Swap
  Selection and Execution}.
\newblock


\bibitem[Quetschlich et~al\mbox{.}(2023)]%
        {quetschlich2023mqtbench}
\bibfield{author}{\bibinfo{person}{Nils Quetschlich}, \bibinfo{person}{Lukas
  Burgholzer}, {and} \bibinfo{person}{Robert Wille}.}
  \bibinfo{year}{2023}\natexlab{}.
\newblock \showarticletitle{{{MQT Bench}}: {Benchmarking Software and Design
  Automation Tools for Quantum Computing}}.
\newblock \bibinfo{journal}{\emph{{Quantum}}} (\bibinfo{year}{2023}).
\newblock


\bibitem[Li et~al\mbox{.}(2023)]%
        {li2020qasmbench}
\bibfield{author}{\bibinfo{person}{Ang Li}, \bibinfo{person}{Samuel Stein},
  \bibinfo{person}{Sriram Krishnamoorthy}, {and} \bibinfo{person}{James Ang}.}
  \bibinfo{year}{2023}\natexlab{}.
\newblock \showarticletitle{Qasmbench: A low-level quantum benchmark suite for
  nisq evaluation and simulation}.
\newblock \bibinfo{journal}{\emph{ACM Transactions on Quantum Computing}}
  \bibinfo{volume}{4}, \bibinfo{number}{2} (\bibinfo{year}{2023}),
  \bibinfo{pages}{1--26}.
\newblock


\bibitem[{McKinney} et~al\mbox{.}(2023)]%
        {2023APSMARG70003M}
\bibfield{author}{\bibinfo{person}{Evan {McKinney}}, \bibinfo{person}{Chao
  {Zhou}}, \bibinfo{person}{Mingkang {Xia}}, \bibinfo{person}{Michael
  {Hatridge}}, {and} \bibinfo{person}{Alex {Jones}}.}
  \bibinfo{year}{2023}\natexlab{}.
\newblock \showarticletitle{{Using Quantum Hardware Speed Limits to Improve
  Basis Gate Selection}}. In \bibinfo{booktitle}{\emph{APS March Meeting
  Abstracts}} \emph{(\bibinfo{series}{APS Meeting Abstracts},
  Vol.~\bibinfo{volume}{2023})}. Article \bibinfo{articleno}{G70.003},
  \bibinfo{numpages}{G70.003}~pages.
\newblock


\end{thebibliography}


\end{document}